\documentclass[lettersize,journal]{IEEEtran}
\usepackage{amsmath,amsfonts}
\usepackage{algorithmic}
\usepackage{algorithm}
\usepackage{array}
\usepackage[caption=false,font=normalsize,labelfont=sf,textfont=sf]{subfig}
\usepackage{textcomp}
\usepackage{stfloats}
\usepackage[hyphens]{url}

\usepackage{verbatim}
\usepackage{graphicx}
\usepackage{cite}
\usepackage[most]{tcolorbox}
\usepackage{layouts}
\usepackage{multirow}
\usepackage{siunitx}
\newcommand\mr[2]{\multirow{#1}{*}{#2}} % handy shortcut macro

\usepackage{bm}
\usepackage{siunitx}
\newcommand\mc[1]{\multicolumn{1}{c}{#1}} % handy shortcut macro
\usepackage{booktabs}

\usepackage{xargs}                      % Use more than one optional parameter in a new commands

\usepackage[colorinlistoftodos,prependcaption,textsize=tiny]{todonotes}
\newcommandx{\atanas}[2][1=]{\todo[linecolor=blue,backgroundcolor=blue!25,bordercolor=blue,#1]{#2}}
\newcommandx{\burak}[2][1=]{\todo[linecolor=green,backgroundcolor=green!25,bordercolor=green,#1]{#2}}
\newcommandx{\panqi}[2][1=]{\todo[linecolor=yellow,backgroundcolor=yellow!25,bordercolor=yellow,#1]{#2}}

\newcommandx{\topic}[2][1=]{\todo[linecolor=red,backgroundcolor=red!25,bordercolor=red,#1]{#2}}
\usepackage{threeparttable}
\usepackage{blindtext}
\usepackage{hyperref}
\hypersetup{hidelinks}
\hypersetup{
	colorlinks   = true, %Colours links instead of ugly boxes
}
\usepackage[capitalize]{cleveref}
\usepackage{orcidlink}
\newcommand{\emailaddress}{burakhan.koyuncu@tum.de}
\usepackage{tikz}
\definecolor{myorange}{RGB}{181, 115, 91}
\definecolor{myblue}{RGB}{125,161,177}
\definecolor{mygreen}{RGB}{151, 183, 150}
\definecolor{mygray}{RGB}{156,156,156}
\DeclareRobustCommand{\drawpatternColor}[1]{
	$\left(\begin{tikzpicture}[x=1.5ex,y=1.5ex,baseline=0.2ex]\draw[color=black,semithick,fill=#1] (0,0) rectangle (1, 1);
	\end{tikzpicture}\right)$}
\usepackage{pifont}% http://ctan.org/pkg/pifont
\newcommand{\cmark}{\ding{51}}%
\newcommand{\xmark}{\ding{55}}%
\makeatletter
\newcommand\thefontsize{The current font size is: \f@size pt}
\begin{document}
	
\title{Efficient Contextformer: Spatio-Channel Window Attention for Fast Context Modeling
	in Learned Image Compression}

\author{A. Burakhan Koyuncu\orcidlink{0000-0002-6291-3476}, \and Panqi Jia\orcidlink{0009-0002-5480-0137}, \and Atanas Boev\orcidlink{0000-0002-6291-3476}, \and Elena Alshina\orcidlink{0000-0001-7099-5371}, \and Eckehard Steinbach\orcidlink{0000-0001-8853-2703},~\IEEEmembership{Fellow,~IEEE}
	% <-this % stops a space
	\thanks{Received April 19, 2023; First Revision received November 6, 2023; Second Revision received December 18, 2023; Accepted February 21, 2024}
	\thanks{A. Burakhan Koyuncu and Eckehard Steinbach are with the Technical University of Munich, School of Computation, Information and Technology, Department of Computer Engineering, Chair of Media Technology, Munich 80333, Germany (e-mail: \href{mailto:\emailaddress}{\emailaddress}).
		
		Panqi Jia is with the Friedrich-Alexander University, Department of Electrical-Electronic-Communication Engineering, Chair of Multimedia Communications and Signal Processing, Erlangen 91058, Germany.
		
		A. Burakhan Koyuncu, Panqi Jia, Atanas Boev and Elena Alshina are with the Huawei Munich Research Center, Munich 80992, Germany.}% <-this % stops a space
}

% The paper headers
\markboth{ACCEPTED FOR IEEE TRANSACTIONS ON CIRCUITS AND SYSTEMS FOR VIDEO TECHNOLOGY}%
{ACCEPTED FOR IEEE TRANSACTIONS ON CIRCUITS AND SYSTEMS FOR VIDEO TECHNOLOGY}

% \IEEEpubid{0000--0000/00\$00.00~\copyright~2021 IEEE}
% Remember, if you use this you must call \IEEEpubidadjcol in the second
% column for its text to clear the IEEEpubid mark.
\IEEEpubid{\begin{minipage}{\textwidth}\ \\[6pt] Copyright~\copyright~2024 IEEE. Personal use of this material is permitted. However, permission to use this material for any other purposes must be obtained from the IEEE by sending an email to pubs-permissions@ieee.org\end{minipage}}
\maketitle

\begin{abstract}
	Entropy estimation is essential for the performance of learned image compression. It has been demonstrated that a transformer-based entropy model is of critical importance for achieving a high compression ratio, however, at the expense of a significant computational effort. In this work, we introduce the Efficient Contextformer (eContextformer) -- a computationally efficient transformer-based autoregressive context model for learned image compression. The eContextformer efficiently fuses the patch-wise, checkered, and channel-wise grouping techniques for parallel context modeling, and introduces a shifted window spatio-channel attention mechanism. We explore better training strategies and architectural designs and introduce additional complexity optimizations. During decoding, the proposed optimization techniques dynamically scale the attention span and cache the previous attention computations, drastically reducing the model and runtime complexity. Compared to the non-parallel approach, our proposal has $\bm{\sim}$145x lower model complexity and $\bm{\sim}$210x faster decoding speed, and achieves higher average bit savings on Kodak, CLIC2020, and Tecnick datasets. Additionally, the low complexity of our context model enables online rate-distortion algorithms, which further improve the compression performance. We achieve up to 17\% bitrate savings over the intra coding of Versatile Video Coding (VVC) Test Model (VTM) 16.2 and surpass various learning-based compression models.
	
	% Entropy estimation is critically important for the performance of learned image compression. It has been demonstrated that using a transformer-based entropy model is greatly beneficial for the compression ratio, however, at the expense of a large computational effort. In this work, we introduce the eContextformer - a computationally efficient transformer to be used for context modelling in learned image compression. It uses a combination of parallel processing, spatio-channel attention, and checkerboard-like grouping of the lantent elements to be processed. Our model features improved decoding speed, lower model complexity and better rate-distortion performance than previous work on the topic. Compared to non-parallel approach, our proposal has ~145x lower model complexity, ~210x lower decoding speed on Kodak, CLIC2020 and Tecnick datasets. Compared to the Versatile Video Coding (VVC) Test Model (VTM) 16.2, the proposed model provides up to 17.1\% bitrate savings and surpasses various learning-based compression models.
\end{abstract}


\section{Introduction}
\IEEEPARstart{O}{nline} media consumption generates an ever-growing demand for higher-quality and lower-bitrate content~\cite{ma2019image,birman2020overview}. Such demand drives advancements in both classical and learned image compression (LIC) algorithms. The early LIC algorithms~\cite{balle2017end,balle2018variational,minnen2018joint,liu2019non,zhou2019multi,cui2021asymmetric,lee2018context,mentzer2018conditional,cheng2020learned,qian2020learning,9067005,koyuncu2021parallel,he2021checkerboard,minnen2020channel,qian2021entroformer} provide competitive rate-distortion performance to classical standards such as JPEG~\cite{wallace1992jpeg}, JPEG2000~\cite{skodras2001jpeg} and BPG~\cite{bellard2015bpg} (intra coding of HEVC~\cite{sze2014high}), and the recent LIC proposals~\cite{he2022elic,guo2021causal,koyuncu2022contextformer} already outperform intra coding of the state-of-the-art video coding standards such as VVC~\cite{ohm2018versatile}.

The best performing LIC frameworks~\cite{balle2017end,balle2018variational,minnen2018joint,liu2019non,zhou2019multi,cui2021asymmetric,lee2018context,mentzer2018conditional,cheng2020learned,qian2020learning,9067005,koyuncu2021parallel,he2021checkerboard,minnen2020channel,qian2021entroformer,guo2021causal,he2022elic,koyuncu2022contextformer,liu2023learned} use an autoencoder, which applies a non-linear, energy compacting transform~\cite{balle2020nonlinear} to an input image. Such frameworks learn a so-called \emph{forward--backward adaptive entropy model}, creating a low-entropy latent space representation of the image. The model estimates the probability distribution by conditioning the distribution to two types of information -- one is a signaled prior (a.k.a. \emph{forward adaptation}~\cite{balle2018variational}), and the other is implicit contextual information extracted by an autoregressive context model (a.k.a. \emph{backward adaptation}~\cite{minnen2018joint}). The performance of the backward adaptation is a major factor in the efficiency of the LIC framework, and methods to improve its performance have been actively investigated. Various context model architectures have been proposed -- 2D masked convolutions using local dependencies~\cite{minnen2018joint,cheng2020learned,lee2018context,liu2019non,cui2021asymmetric,zhou2019multi}; channel-wise autoregressive mechanisms exploiting channel dependencies~\cite{mentzer2018conditional,minnen2020channel,koyuncu2022contextformer}; non-local simplified attention-based models capturing long-range spatial dependencies~\cite{guo2021causal,qian2020learning}; sophisticated transformer-based models leveraging a content-adaptive context modeling~\cite{qian2021entroformer,koyuncu2022contextformer,liu2023learned}.

Using an autoregressive model requires recursive access to previously processed data, which results in slow decoding time and inefficient utilization of NPU/GPU. To remedy this, some
researchers proposed to optimize the decoding process by using wavefront parallel processing (WPP)~\cite{9067005,cheng2020learned,koyuncu2022contextformer} inspired by a similar approach used in classical codecs~\cite{chi2012parallel}. While using WPP significantly increases parallelism, it still requires a large number of autoregressive steps for processing large images. A more efficient approach is to split the latent elements into groups and code each group separately~\cite{koyuncu2021parallel,he2021checkerboard,minnen2020channel,qian2021entroformer,he2022elic}. The latent elements might be split into, e.g., spatial patches~\cite{koyuncu2021parallel}, channel segments~\cite{minnen2020channel}, or using a checkered pattern~\cite{he2021checkerboard}. A combination of the channel-wise and the checkered grouping has also been studied~\cite{he2022elic}. Such parallelization approaches can reduce decoding time by 2-3 orders of magnitude at the cost of higher model complexity and up to a 3\% performance drop. For instance, the patch-wise model~\cite{koyuncu2021parallel} uses a recurrent neural network (RNN) to share information between patches. Other works~\cite{minnen2020channel,he2022elic,liu2023learned} use a channel-wise model and implement additional convolutional layers to combine decoded channel segments.\IEEEpubidadjcol

Our previous work proposed a transformer-based entropy model with spatio-channel attention (Contextformer)~\cite{koyuncu2022contextformer}. In the same work, we presented a framework using our entropy model, which outperforms contemporary LIC frameworks~\cite{balle2017end,balle2018variational,minnen2018joint,liu2019non,zhou2019multi,cui2021asymmetric,lee2018context,mentzer2018conditional,cheng2020learned,qian2020learning,minnen2020channel,qian2021entroformer,guo2021causal}. However, the model complexity of Contextformer makes it unsuitable for real-time operation. In this work, we propose a fast and low-complexity version of the Contextformer, which we call Efficient Contextformer (eContextformer). We extend the window attention of~\cite{liu2021swin} to spatio-channel window attention in order to achieve a high-performance and low-complexity context model. Additionally, we use the checkered grouping to increase parallelization and reduce autoregressive steps in context modeling. By exploiting the properties of eContextformer, we also propose algorithmic optimizations to reduce complexity and runtime even further.

In terms of PSNR, our model surpasses VTM 16.2 \cite{jvet2019versatile} (intra)\footnote{Throughout this work, we use VTM~16.2~\cite{jvet2019versatile} in intra coding mode, i.e., with configuration file of \textit{encoder\_intra\_vtm.cfg} and input format of YUV444.} on Kodak \cite{franzen1999kodak}, CLIC2020~\cite{CLIC2020} (Professional and Mobile), and Tecnick \cite{asuni2014testimages} datasets, providing average bitrate savings of 10.9\%, 12.3\%, 6.9\%, and 13.2\%, respectively. Our optimized implementation requires 145x less multiply-accumulate (MAC) operations and takes 210x less decoding time than the earlier Contextformer. It also contains significantly fewer model parameters compared to other channel-wise autoregressive and transformer-based prior-art context models~\cite{minnen2020channel,qian2021entroformer,liu2023learned}. Furthermore, due to its manageable complexity, eContextformer can be used with an online rate-distortion optimization (oRDO) similar to the one in~\cite{campos2019content}. With oRDO enabled, the performance of our model reaches up to 17\% average bitrate saving over VTM~16.2~\cite{jvet2019versatile}.

In the next section we overview of the related work and introduce the necessary terminology. In Section~III, we introduce our proposal for an efficient and high-performance transformer-based context model. In Section~IV, we describe the experiments and the experimental results, and in Section~V, we present the conclusions.
\begin{figure*}[!t]
\centering
% \captionsetup[subfigure]{labelformat=empty}
\tcbset{rounded corners,colframe=black,left=0mm,right=0mm,top=0mm,bottom=0mm,boxsep=0mm,arc=0mm,boxrule=1pt}
\subfloat[]{\includegraphics[page=9,width=0.165\textwidth,trim={1.94cm 12.4cm 2.5cm 2.25cm},clip]{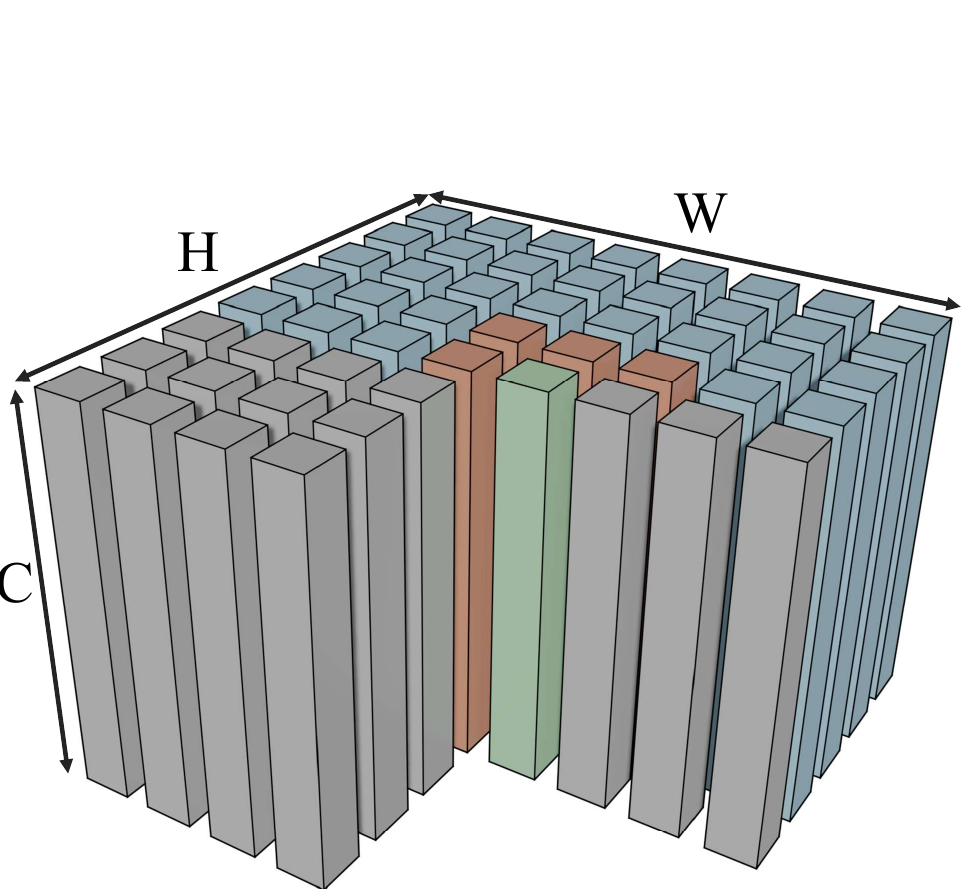}}
\hfill
    \subfloat[]{\includegraphics[page=10,width=0.165\textwidth,trim={1.94cm 12.4cm 2.5cm 2.55cm},clip]{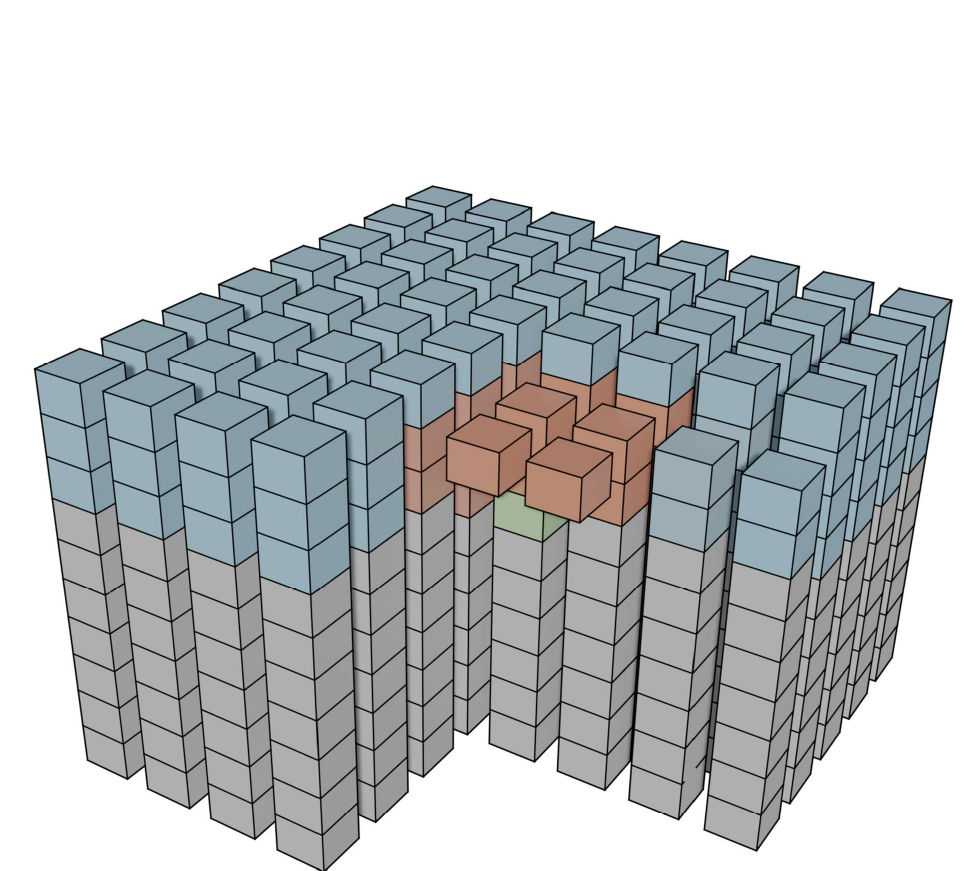}}
\hfill
    \subfloat[]{\includegraphics[page=11,width=0.165\textwidth,trim={1.94cm 12.4cm 2.5cm 2.55cm},clip]{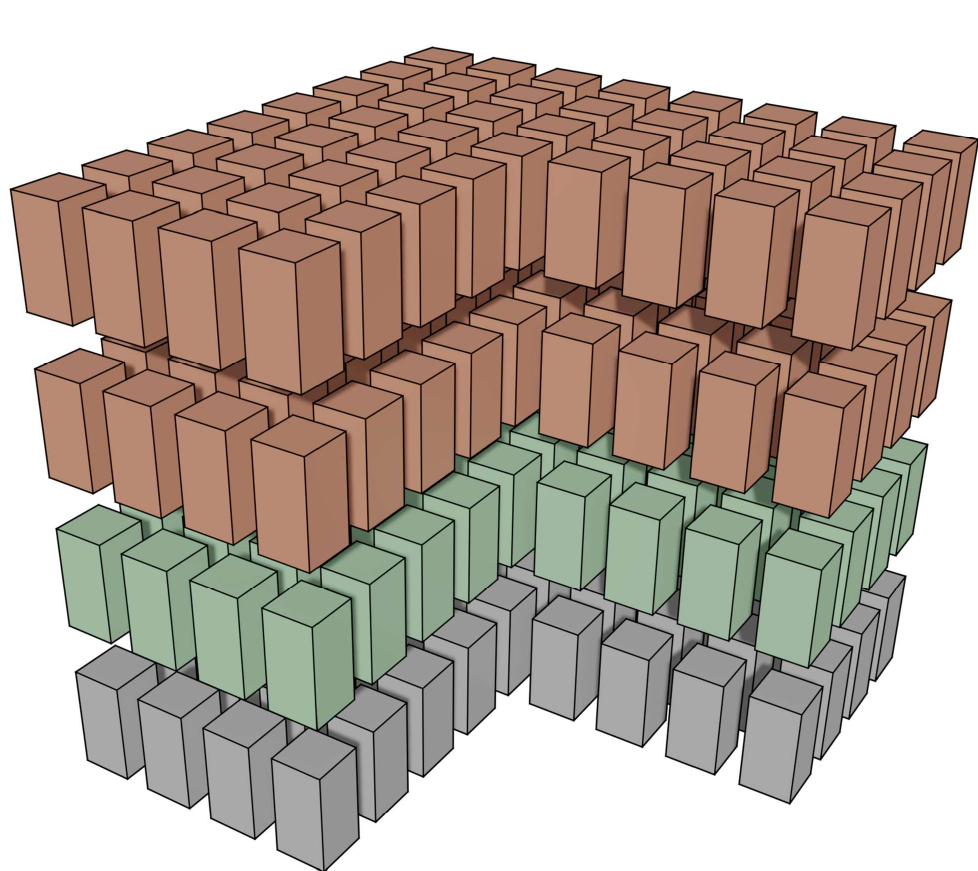}}
% \hfill
%     \subfloat[]{\includegraphics[page=12,width=0.165\textwidth,trim={1.94cm 12.4cm 2.5cm 2.55cm},clip]{figures/model7.pdf}}
\hfill
    \subfloat[]{\includegraphics[page=13,width=0.165\textwidth,trim={1.94cm 12.4cm 2.5cm 2.55cm},clip]{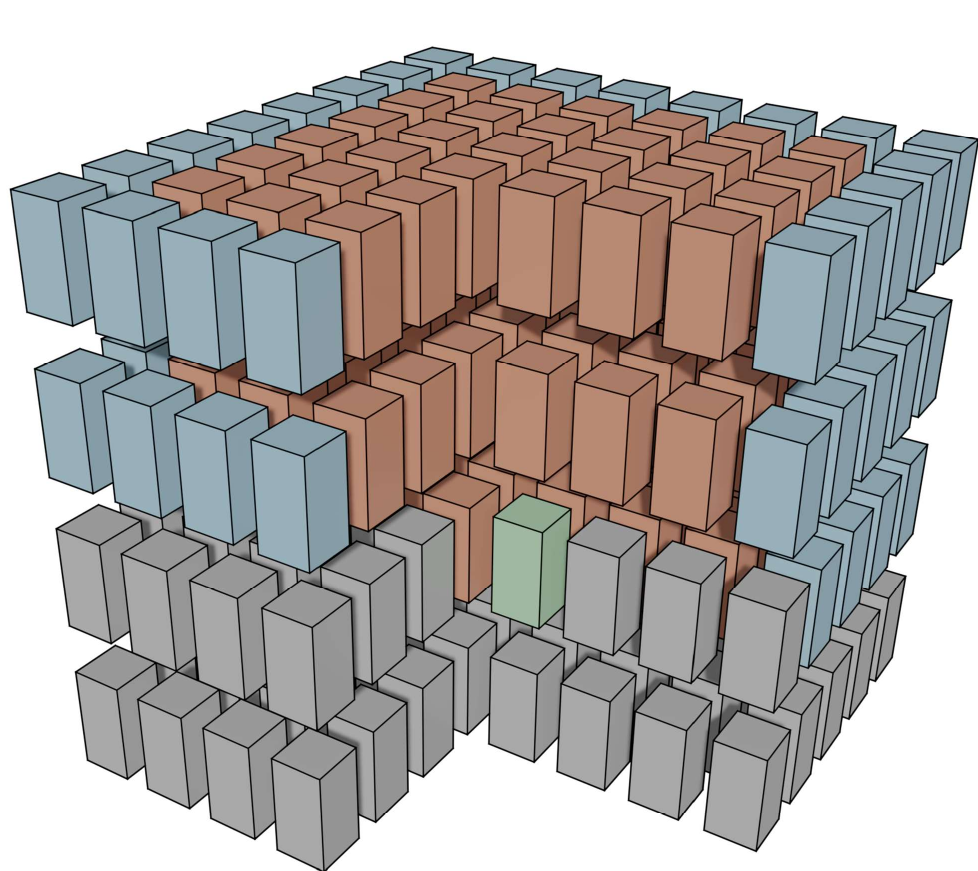}}
\hfill
    \subfloat[]{\includegraphics[page=14,width=0.165\textwidth,trim={1.94cm 12.4cm 2.5cm 2.55cm},clip]{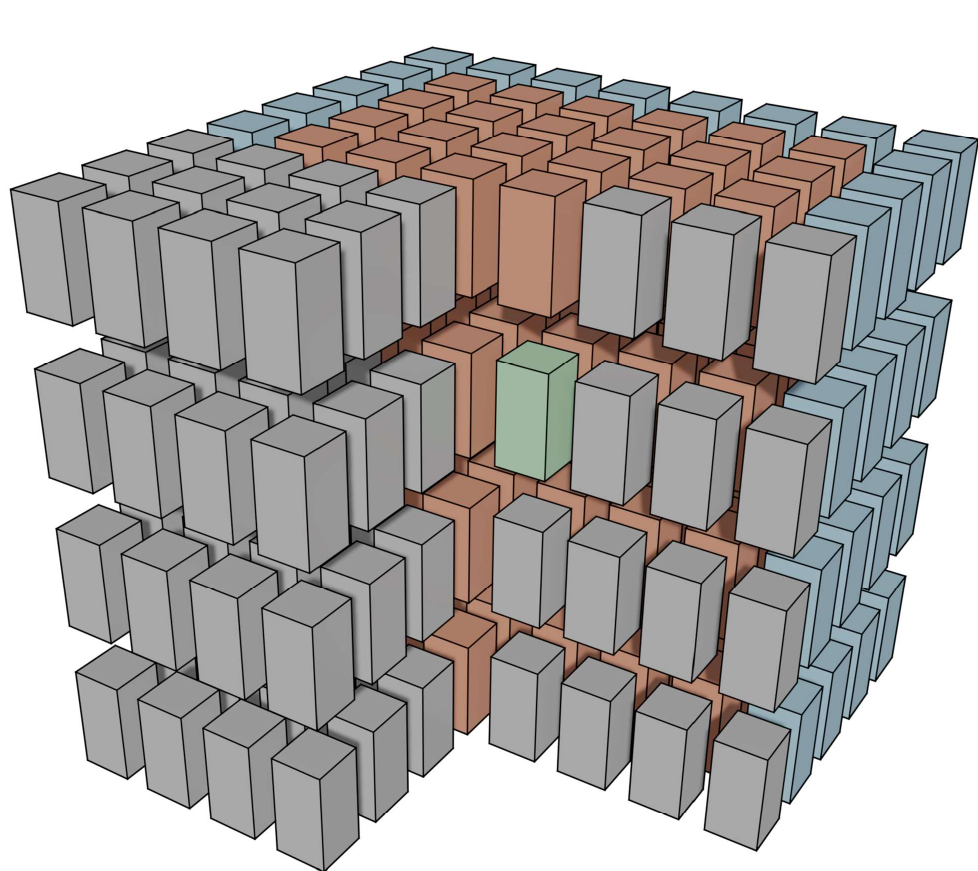}}
\caption{Illustration of the context modeling process, where the symbol probability of the current latent variable {\drawpatternColor{mygreen}} estimated by aggregating the information of the latent variables {\drawpatternColor{myorange}}. The previously decoded latent elements not joining to context modeling and yet to be coded elements are depicted as {\drawpatternColor{myblue}} and {\drawpatternColor{mygray}}, respectively. The illustrated context models are (a) the model with 2D masked convolutions~\cite{minnen2018joint,cheng2020learned}, (b) the model with 3D masked convolutions~\cite{liu2019non,mentzer2018conditional}, (c) channel-wise autoregressive model~\cite{minnen2020channel}, and (d\textendash e) Contextformer with sfo and cfo coding mode~\cite{koyuncu2022contextformer}, respectively.}
\label{fig_ctx_models1}
\end{figure*}

\section{Related Work}
\subsection{Transformers in Computer Vision}
\label{sec_1}
The underlying principle of the transformer network~\cite{vaswani2017attention} can be summarized as a learned projection of sequential vector embeddings \mbox{$\bm{x} \in\mathbb{R}^{S\times d_e}$} into sub-representations of query \mbox{$\bm{Q} \in\mathbb{R}^{S\times d_e}$}, key $\bm{K}\in\mathbb{R}^{S\times d_e}$ and value $\bm{V}\in\mathbb{R}^{S\times d_e}$, where $S$ and $d_e$ denote the sequence length and the embedding size, respectively. Subsequently, a scaled dot-product calculates the attention, which weighs the interaction between the query and key-value pairs. The $\bm{Q}$, $\bm{K}$ and $\bm{V}$ are split into $h$ groups, known as \emph{heads}, which enables parallel computation and greater attention granularity. The separate attention of each head is finally combined by a learned projection $W$. After the attention module, a point-wise multi-layer perceptron (MLP) is independently applied to all positions in $S$ to provide non-linearity. For autoregressive tasks, \emph{attention masking} is required to ensure the causality of the interactions within the input sequence. The multi-head attention with masking can be described as:
\begin{subequations}
	\begin{align}
		&Attn(\bm{Q},\bm{K},\bm{V})= \text{concat}({head}_1,\ldots,{head}_h)\bm{W},\\
		&head_i(\bm{Q}_i,\bm{K}_i,\bm{V}_i)=\text{softmax}\left(\frac{\bm{Q}_i\bm{K}_i^T}{d_k}\odot\bm{M}\right)\bm{V}_i,
		\label{eq:multihead}
	\end{align}
\end{subequations}
where the mask $\bm{M} \in\mathbb{R}^{S\times S}$ contains weights of $-\infty$ in the positions corresponding to interactions with elements which are not processed yet, and ones for the rest. The operator $\odot$ denotes the Hadamard product.

Once trained, a typical image processing algorithm that uses a convolutional neural network (CNN)~\cite{NIPS2012_c399862d} uses static weights to calculate its output. In contrast, the attention mechanism in transformers uses content-adaptive weighting, which enables a transformer to capture long-distance relations~\cite{naseer2021intriguing}. This feature makes transformers suitable for a variety of computer vision tasks~\cite{carion2020end,dosovitskiy2020image,jiang2021transgan,esser2021taming}. As a downside, the computational complexity of a vanilla transformer grows quadratically with the length of the sequence $S$. One solution to the complexity growth is the Vision Transformer (ViT) proposed by Dosovitskiy et al.~\cite{dosovitskiy2020image}. The ViT decomposes the input into non-overlapping 2D patches and computes global attention between the patches. Alternatively, \cite{parmar2018image,esser2021taming,koyuncu2022contextformer} propose using sliding window attention to limit the attention span, as computing local attention has lower complexity than computing the global one. Another solution is the Hierarchical Vision Transformer using Shifted Windows, also called \textit{Swin transformer}, proposed in Liu et al.~\cite{liu2021swin}. The attention is computed in non-overlapping windows and is later shared between the windows by cyclically shifting the window positions.

\subsection{Learned Image Compression}
In~\cite{balle2020nonlinear}, the authors propose a learned non-linear transform coding for lossy image compression. They use an autoencoder with an entropy-constrained bottleneck, with the goal of achieving high fidelity of the transform while reducing the symbol entropy. The encoder uses an analysis transform $g_s$ to project the input image $\bm{x}$ to a latent variable $\bm{y}$ with lower dimensionality. The result is quantized to $\bm{\hat{y}}$ by a quantization function $Q$ and encoded into a bitstream using a lossless codec such as~\cite{rissanen1979arithmetic}. The decoder reads the bitstream of $\bm{\hat{y}}$ and applies a synthesis transform $g_s$ to output the reconstructed image $\bm{\hat{x}}$.
% 	p_{\bm{\hat{y}_{i}}}(\bm{\hat{y}_{i}}|\bm{\hat{z}}) &\leftarrow  g_{e}( g_{c}(\bm{\hat{y}_{<i}}; \bm{\theta_{c}}), &h_s(\bm{\hat{z}}; \bm{\theta_s}); &\bm{\psi_{e}}),
The entropy of $\bm{\hat{y}}$ is minimized by learning its probability distribution. Entropy modeling is built using two methods, backward and forward adaptation~\cite{balle2020nonlinear}. 
The forward adaption uses a hyperprior estimator, i.e., a second autoencoder with its own analysis transform $h_a$ and synthesis transform $h_s$. The hyperprior creates a side, separately encoded channel $\bm{\hat{z}}$. A factorized density model~\cite{balle2017end} learns local histograms to estimate the probability mass $p_{\bm{\hat{z}}}(\bm{\hat{z}}|\bm{\psi_f})$ with model parameters $\bm{\psi_f}$. The backward adaptation uses a context model $g_{cm}$, which estimates the entropy of the current latent element $\bm{\hat{y}_i}$ using the previously coded elements $\bm{\hat{y}_{<i}}$ in an autoregressive fashion. The entropy parameters network $g_{e}$ uses the outputs of the context model and the hyperprior to parameterize the conditional distribution $p_{\bm{\hat{y}}}(\bm{\hat{y}}|\bm{\hat{z}})$. The lossy LIC framework can be formulated as:
\begin{subequations}
	\begin{align}
		\bm{\hat{y}}&= Q(g_a(\bm{x}; \bm{\phi_a})),\\
		\bm{\hat{x}}&= g_s(\bm{\hat{y}}; \bm{\phi_s})),\\
		\bm{\hat{z}}&= Q(h_a(\bm{\hat{y}}; \bm{\theta_a})),\\
		p_{\bm{\hat{y}_{i}}}(\bm{\hat{y}_{i}}|\bm{\hat{z}})&\leftarrow g_{e}( g_{c}(\bm{\hat{y}_{<i}}; \bm{\theta_{c}}), h_s(\bm{\hat{z}}; \bm{\theta_s}); \bm{\psi_{e}}),
		\label{eq:1}
	\end{align}
\end{subequations}
with the loss function $\mathcal{L}$ of end-to-end training:
\begin{subequations}
	\begin{align}
		\mathcal{L}(\bm{\phi}, \bm{\theta}, \bm{\psi}) &= \mathbf{R}(\bm{\hat{y}}) + \mathbf{R}(\bm{\hat{z}}) + \lambda \cdot \mathbf{D}(\bm{x}, \bm{\hat{x}}),\label{eq_2}\\
		&= \mathbb{E}[\log_2(p_{\bm{\hat{y}}}(\bm{\hat{y}}|\bm{\hat{z}}))] + \mathbb{E}[\log_2(p_{\bm{\hat{z}}}(\bm{\hat{z}}|\bm{\psi}))]\notag\\
		&\quad + \lambda \cdot \mathbf{D}(\bm{x}, \bm{\hat{x}}),
	\end{align}
\end{subequations}
where $\bm{\phi}$, $\bm{\theta}$ and $\bm{\psi}$ are the optimization parameters of their corresponding transforms. The Lagrange multiplier $\lambda$ regulates the trade-off between distortion $\mathbf{D}(\cdot)$ and rate $\mathbf{R}(\cdot)$.

\begin{figure*}[!t]
\centering
% \captionsetup[subfigure]{labelformat=empty}
\includegraphics[width=\textwidth]{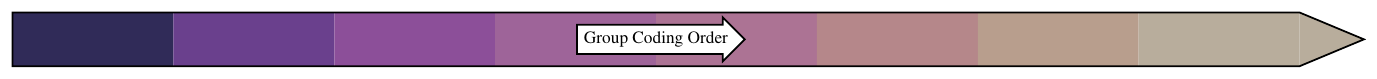}\vspace{-0.9\baselineskip}
\subfloat[]{\includegraphics[page=15,width=0.165\textwidth,trim={2.25cm 8.8cm 2.5cm 4.8cm},clip]{figures/model7.pdf}}
\hfill
\subfloat[]{\includegraphics[page=16,width=0.165\textwidth,trim={2.25cm 11.4cm 2.5cm 2.25cm},clip]{figures/model7.pdf}}
\hfill
\subfloat[]{\includegraphics[page=17,width=0.165\textwidth,trim={2.25cm 11.4cm 2.5cm 2.25cm},clip]{figures/model7.pdf}}
\hfill
\subfloat[\label{fig_ctx_models2:4}]{\includegraphics[page=18,width=0.165\textwidth,trim={2.25cm 11.4cm 2.5cm 2.25cm},clip]{figures/model7.pdf}}
\hfill
\subfloat[\label{fig_ctx_models2:5}]{\includegraphics[page=19,width=0.165\textwidth,trim={2.25cm 11.4cm 2.5cm 2.25cm},clip]{figures/model7.pdf}}
\caption{Illustration of different parallelization techniques for the context modeling in (a) patch-wise grouping~\cite{koyuncu2021parallel}, (b) checkered grouping~\cite{he2021checkerboard}, (c) channel-wise grouping~\cite{minnen2020channel}, and (d-e) combination of checkered and channel-wise grouping with sfo and cfo coding, respectively. All latent elements within the same group (depicted with the same color) are coded simultaneously, while the context model aggregates the information from the previously coded groups. For instance, \cite{koyuncu2021parallel,he2021checkerboard} use 2D masked convolutions in the context model, and \cite{minnen2020channel} applies multiple CNNs to channel-wise concatenated groups. The context model of~\cite{he2022elic} combines the techniques of \cite{minnen2020channel,he2021checkerboard} and can be illustrated as in (d). Our proposed model (eContextformer), as well as the experimental model (pContextformer), use the parallelization techniques depicted in (d\textendash e). However, our models employ spatio-channel attention in context modeling and do not require additional networks for channel-wise concatenation.}
\label{fig_ctx_models2}
\end{figure*}

\subsection{Efficient Entropy Modeling}
As an autoregressive context model yields a significant increase in coding efficiency, multiple architectures have been proposed (see \cref{fig_ctx_models1}). We categorized those into three groups; (1) exploiting spatial dependencies; (2) exploiting cross-channel dependencies; (3) increased content-adaptivity in the entropy estimation~\cite{koyuncu2022contextformer}.

The first group contains works such as Minnen et al.~\cite{minnen2018joint}, which proposed to use a 2D masked convolution to exploit spatially local relations in the latent space. In order to capture a variety of spatial dependencies, \cite{cui2021asymmetric,zhou2019multi} implemented a multi-scale context model using multiple masked convolutions with different kernel sizes. The second group encompasses context models such as 3D context model~\cite{liu2019non,mentzer2018conditional,tang2022joint} and channel-wise autoregressive context model~\cite{minnen2020channel}. Such models are autoregressive in channel dimension; thus, they can use information derived from previously coded channels. In~\cite{liu2019non,mentzer2018conditional,tang2022joint}, the authors use 3D masked convolutions which exploit both spatial and cross-channel dependencies, while the proposal in~\cite{minnen2020channel} uses cross-channel dependencies only. In the third group, attention-based models provide better content-adaptivity than the CNN-based ones~\cite{guo2021causal,qian2020learning,qian2021entroformer,koyuncu2022contextformer}. Early proposals~\cite{guo2021causal,qian2020learning} combine 2D masked convolutions and simplified attention to capture local and global correlations of latent elements. Recent studies~\cite{qian2020learning,koyuncu2022contextformer} have been investigating transformer-based architectures to dynamically capture long-ranged spatial dependencies with the attention mechanism. Moreover, extending the attention mechanism for exploiting cross-channel dependencies was explored for high-performance context modeling~\cite{koyuncu2022contextformer}.

In the decoder, the context modeling requires an autoregressive processing of the latent elements. Such serial processing results in slower decoding time and lower NPU/GPU efficiency. In \cref{fig_ctx_models2}, we present an overview of techniques that increase parallelization by grouping the elements in various ways~\cite{koyuncu2021parallel,he2021checkerboard,minnen2020channel}. In \cite{koyuncu2021parallel}, we proposed a patch-wise context model processing the latent variable within spatially non-overlapping patches. Inside each patch, a serial context model is applied while the inter-patch relations are aggregated with an RNN. The authors of~\cite{he2021checkerboard} proposed grouping of the latent variables according to a checkered pattern. They use a 2D masked convolution with a checkered pattern in context modeling. Both \cite{koyuncu2021parallel} and \cite{he2021checkerboard} outperform their serial baseline in high bitrates but are inferior to them in low bitrates. The model in~\cite{minnen2020channel} is another example of a parallel context model, which uses channel-wise grouping. He et al.~\cite{he2022elic} combined checkered grouping with channel-wise context modeling. Liu et al.~\cite{liu2023learned} improved the architecture in~\cite{he2022elic} with transformer-based transforms. Both models~\cite{he2022elic,liu2023learned} reached significantly high rate-distortion performance. Wang et al.~\cite{wang2023evc} proposed a checkered grouping with an pattern for the channel dimension in addition to the spatial ones. The model approximates channel-wise grouping with a significantly a lower model complexity but it reaches a lower rate-distortion performance.

The studies~\cite{he2022elic,liu2023learned,wang2023evc,fu2023asymmetric,yang2023computationally,tang2022joint,jia2022learning} investigate better analysis and synthesis transforms to reach high rate-distortion performance for a good complexity trade-off. For instance, in the model of \cite{liu2023learned}, all the transform layers are replaced with transformers, which significantly improve the performance over their CNN-based baseline. The model of~\cite{tang2022joint} employ transforms with graph attention layers, which trades higher performance for higher complexity. In~\cite{wang2023evc,fu2023asymmetric,yang2023computationally}, the authors proposed asymmetrical compression frameworks, which has a decoder with orders of magnitude lower complexity than the encoder, and even avoid using context models~\cite{yang2023computationally}. Although the model of~\cite{fu2023asymmetric} reaches a lower model complexity compared to their relative baseline~\cite{cheng2020learned}, it still has a slow decoding speed due to their 2D serial context model. In general, asymmetric frameworks aim to reach ultra-low complexities, but their rate-distortion performance is not competitive with higher complexity models.

\subsection{Online Rate-Distortion Optimization}
Training an LIC encoder on a large dataset results in a globally optimal but locally sub-optimal network parameters~$(\bm{\phi}, \bm{\theta}, \bm{\psi})$. For a given input image $x$, a more optimal set of parameters might exist. This problem is known as the \emph{amortization gap}~\cite{yang2020improving,gao2022flexible,campos2019content}. Producing optimal network parameters for each $x$ is not feasible during encoding due to the signaling overhead and computational complexity involved. In \cite{yang2020improving,campos2019content}, the authors proposed an online rate-distortion optimization (oRDO), which adapts $\bm{y}$ and $\bm{z}$ during encoding. The oRDO algorithms first initialize the latent variables by passing the input image $\bm{x}$ through the encoder without tracking the gradients of the transforms. Then, the network parameters are kept frozen, and the latent variables are iteratively optimized using (\ref{eq_2}). Finally, the optimized variables $(\bm{\hat{y}_{opt}},\bm{\hat{z}_{opt}})$ are encoded into the bitstream. The oRDO does not influence the complexity or the runtime of the decoder, but it significantly increases the encoder complexity. Therefore, oRDO is suitable for LIC frameworks with low complexity entropy models~\cite{balle2018variational,minnen2018joint,cheng2020learned}.

\section{Efficient Contextformer}
\subsection{Contextformer}
\label{sec_2a}
In a previous work, we proposed a high-performance transformer-based context model (a.k.a Contextformer)~\cite{koyuncu2022contextformer}. Compared to the prior transformer-based context modeling method~\cite{qian2021entroformer}, Contextformer introduces additional autoregression for the channel dimension by splitting the latent tensor into several channel segments $N_{cs}$. This enables adaptively exploiting both spatial and channel-wise dependencies in the latent space with a spatio-channel attention mechanism. Unlike~\cite{minnen2020channel}, each segment is processed by the same model without requiring additional complex architectures for aggregating channel segments. Contextformer architecture is based on ViT~\cite{dosovitskiy2020image}, but the patch-wise global attention mechanism is replaced with a sliding window attention similar to the one used in~\cite{esser2021taming}. The receptive field traverses the spatial dimensions and grows in the channel dimension. The proposed attention mechanism has two different coding modes -- \textit{spatial-first-order} (sfo) and \textit{channel-first-order} (cfo). The coding modes differ in the order of the autoregressive steps within the sliding window, resulting in a prioritization of the correlations in the primarily processed dimension (see \cref{fig_ctx_models2:4,fig_ctx_models2:5}). Although the Contextformer provides a significant rate-distortion performance, it is infeasible for real-time operation due to the serial processing involved.

\subsection{Exploring Parallelization Techniques}
\label{sec_2}
We analyzed the effect of state-of-the-art parallelization techniques on the Contextformer. We observed a few issues that can be improved: (1) overly large kernel size of the attention mechanism; (2) a discrepancy between train and test time behavior; (3) a large number of autoregressive steps. For an input image $\bm{\hat{x}}\in\mathbb{R}^{H \times W \times 3}$ with a latent representation $\bm{\hat{y}}\in\mathbb{R}^{\frac{H}{16} \times \frac{W}{16} \times C}$ (with height $H$, width $W$ and the number of channels $C$), the complexity $\mathbb{C}$ of the proposed attention mechanism expressed in number of MAC operations is:
\begin{subequations}
	\begin{align}
		\mathbb{C}_{\text{attn}} &= N_{win} (2N^2 d_e + 4N {d_e}^2),\\
		N_{win} &= \frac{HWN_{cs}}{256s^2},\quad N= K^2 N_{cs},
		\label{eq:complexity}
	\end{align}
\end{subequations}
where the spatial kernel size $K$ and the number of channel segments $N_{cs}$ defines the sequence length $N$ inside a spatio-channel window. The window operation is performed $N_{win}$ times with a stride of $s{=}1$ over $\bm{\hat{y}}$.

The Contextformer is trained with a large spatial kernel ($K{=}16$) on $256{\times}256$ image crops. Therefore, the model learns a global attention mechanism. During inference, it uses sliding window attention to reduce computational complexity. Compared to global attention used in~\cite{qian2020learning}, the sliding window attention provides lower latency. However, the Contextformer does not learn the sliding behavior, which creates a discrepancy between training and inference results. This problem could be fixed by training with sliding window attention and either increasing the image crops and keeping the kernel size the same, or by decreasing the kernel size and keeping the crop size intact. Training with larger crop size does not affect the complexity and still increases the compression performance. In this case, direct comparison with prior art trained on small crops would be unfair and would make the effects of parallelization less noticeable. Training with a smaller kernel provides biquadratic complexity reduction for each window. Moreover, the Contextformer requires $N_{win}$ autoregressive steps, which cannot be efficiently implemented on a GPU/NPU. Based on those observations, we built a set of parallel Contextformers (pContextformer), which fuse the patch-wise context~\cite{koyuncu2021parallel} and checkerboard context model~\cite{he2021checkerboard} and need eight autoregressive steps for $N_{cs}{=}4$. 
\begin{figure}[!t]
\centering
\tcbset{rounded corners,colframe=black,left=0mm,right=0mm,top=0mm,bottom=0mm,boxsep=0mm,arc=0mm,boxrule=1pt}
\includegraphics[width=1.0\linewidth]{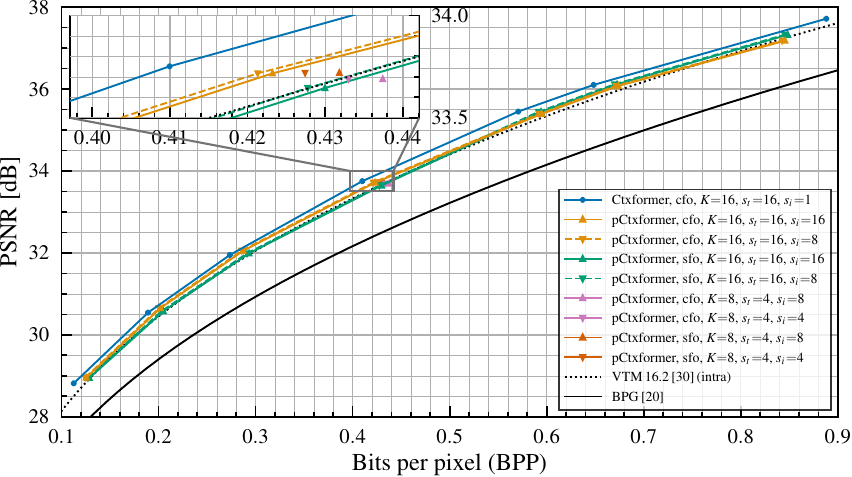}
\caption{Experimental study on Kodak dataset \cite{franzen1999kodak}, comparing the rate-distortion performance of different model configurations}
\label{fig_abl1}
\end{figure}

We trained the set of models with $256{\times}256$ image crops for ${\sim}65\%$ of the number of iterations used for training in \cite{koyuncu2022contextformer}. We varied the coding method (sfo or cfo), the kernel size $K$, and the stride during training $s_t$ and inference $s_i$. The stride defines whether the windows are processed in overlapping or separately. To achieve the overlapping window process, we set the stride to $\frac{K}{2}$. For $K{=}16$, the model could learn only global attention, which could be replaced with a local one during the inference. Following the methodology of JPEG-AI~\cite{ascenso2023jpeg}, we measured the performance of the codec in Bjøntegaard Delta rate (BD-Rate)~\cite{bjontegaard2001calculation} over VTM 16.2~\cite{jvet2019versatile}, and the complexity of each context model in kilo MAC operations per pixel (kMAC/px). For simplicity, we measured the complexity of a single pass on the encoder side, where the whole latent variable $\bm{\hat{y}}$ is processed at once with the given causality mask. We present the results in \cref{fig_abl1} and \cref{tab_abl1,tab_abl2}) and make the following observations:

\paragraph{When trained with global attention ($s_t{=}K{=}16$)} the spatial-first coding is better than channel-first coding at high bitrates but worse at low bitrates. At low bitrates, the spatial-first coding cannot efficiently exploit spatial dependencies due to the usage of non-overlapping windows on the sparse latent space. Also, spatial-first coding benefits more from the overlapping window attention applied at inference time~($s_i{<}K$).

\paragraph{Trained with overlapping attention windows ($s_t{<}K$)} the spatial-first coding outperforms channel-first coding. Moreover, using overlapping windows at inference time helps the models with a small kernel size ($K{=}8$) to reach performance close to the ones with a larger kernel~($K{=}16$).

\paragraph{In general}  pContextformer models can provide more than 100x complexity reduction for a $3\%$ performance drop. Theoretically, more efficient models are possible with using overlapping window attention at training and inference time. However, the simple overlapping window attention is still sub-optimal since it increases the complexity four-fold.

\subsection{Proposed Model}
Based on previous experiments we propose the \textit{Efficient Contextformer} (eContextformer) as an improved version of our previous architecture~\cite{koyuncu2022contextformer}. It uses the same compression framework as~\cite{koyuncu2022contextformer,cui2021asymmetric}. The analysis transform $g_a$ comprises four $3{\times}3$ convolutional layers with a stride of 2, GDN activation function~\cite{balle2015density}, and a single residual attention module without the non-local mechanism~\cite{cheng2020learned,cui2021asymmetric,zhang2018residual}, same as in our previous work\footnote{In \cite{koyuncu2022contextformer}, we misclassified the attention module as being a non-local one. It uses is the same architecture as in~\cite{zhang2018residual}, but without the non-local module. A similar architecture is proposed in~\cite{cheng2020learned}.}. The synthesis transform $g_s$ closely resembles $g_a$ and implements deconvolutional layers with inverse GDN (iGDN). In order to expand receptive fields and reduce quantization error, $g_s$ utilizes residual blocks~\cite{liu2019non} in its first layer, and an attention module. The entropy model integrates a hyperprior network, universal quantization, and estimates $p_{\bm{\hat{y}_{i}}}(\bm{\hat{y}_{i}}|\bm{\hat{z}})$ with a Gaussian Mixture Model (GMM)~\cite{cheng2020learned} with $k_m{=}3$ mixtures. In contrast to~\cite{koyuncu2022contextformer}, we use $5{\times}5$ convolutions in lieu of the $3{\times}3$ ones in the hyperprior in order to align our architecture with the recent studies~\cite{he2022elic,minnen2020channel}. 

\begin{table}
    % \ra{1.3}
    % \addtolength{\tabcolsep}{-4pt}
    
	\centering
	\caption{Rate savings over VTM~16.2~\cite{jvet2019versatile} (intra) and complexity of various pContextformers with $K{=}16$ compared to~\cite{koyuncu2022contextformer}, showing the effect of coding mode and using overlapped windows during inference}
	\label{tab_abl1}
	\begin{tabular}{@{}lcc|c|c@{}}

		\toprule
		       & Coding &  Ovp. Win. &   &  BD-Rate      \\
		Method & Mode  &    @Inf.   & kMAC/px & [\%] \\
		\midrule
		VTM~16.2~\cite{jvet2019versatile} (intra) & \textendash & \textendash & \textendash & \phantom{$-$}0.0\phantom{$-$} \\
		Contextformer~\cite{koyuncu2022contextformer} & \mr{1}{cfo}    & \cmark &\mr{1}{$36\cdot10^3$}  & {$-6.9$}\phantom{$-$} \\
		\midrule
		\mr{4}{pContextformer} & cfo  &  \xmark  & \phantom{0}320 & {$-3.1$}\phantom{$-$} \\
		                        & cfo   & \cmark  & 1072 & {$-3.5$}\phantom{$-$} \\
		                        & sfo   & \xmark  & \phantom{0}320 & {$-1.2$}\phantom{$-$} \\
		                        & sfo   & \cmark  & 1072 & {$-1.9$}\phantom{$-$} \\

		\bottomrule
	\end{tabular}
	   % \addtolength{\tabcolsep}{-4pt} 
\end{table}
\begin{table}
    % \ra{1.3}
    % \addtolength{\tabcolsep}{4pt}
    
	\centering
	\caption{Performance and complexity of various pContextformers w.r.t. kernel size, coding mode, and using overlapped windows during inference and training}
	\label{tab_abl2}
	\begin{tabular}{@{}cccc|c|c@{}}

		\toprule
		 Coding &         & \multicolumn{2}{c|}{Ovp. Win.} &  &       \\
		 Order  & $K$     &  @Tr. & @Inf.   & kMAC/px & $\Delta$BPP  \\
		\midrule
		 cfo & 8  & \cmark  & \xmark   & \phantom{0}195 & {\phantom{-}0} \\
		 cfo & 8  & \cmark  & \cmark  & \phantom{0}710 & {${\phantom{1}-}4\cdot10^{-3}$} \\
		 sfo & 8  & \cmark  & \xmark  & \phantom{0}195 & {${\phantom{1}-}6\cdot10^{-3}$} \\
		 sfo & 8  & \cmark  & \cmark  & \phantom{0}710 & {$-10\cdot10^{-3}$} \\
		 cfo & 16 & \xmark  & \xmark  & \phantom{0}320 & {$-14\cdot10^{-3}$} \\
		 cfo & 16 & \xmark  & \cmark  & 1072 & {$-16\cdot10^{-3}$} \\
		\bottomrule
	\end{tabular}
	   % \addtolength{\tabcolsep}{-4pt}
\end{table}

\begin{figure*}[!t]
\tcbset{rounded corners,colframe=black,left=0mm,right=0mm,top=0mm,bottom=0mm,boxsep=0mm,arc=0mm,boxrule=1pt}
\centering
\includegraphics[page=1,width=0.85\textwidth,trim={1.7cm 21.25cm 1.7cm 1.2cm},clip]{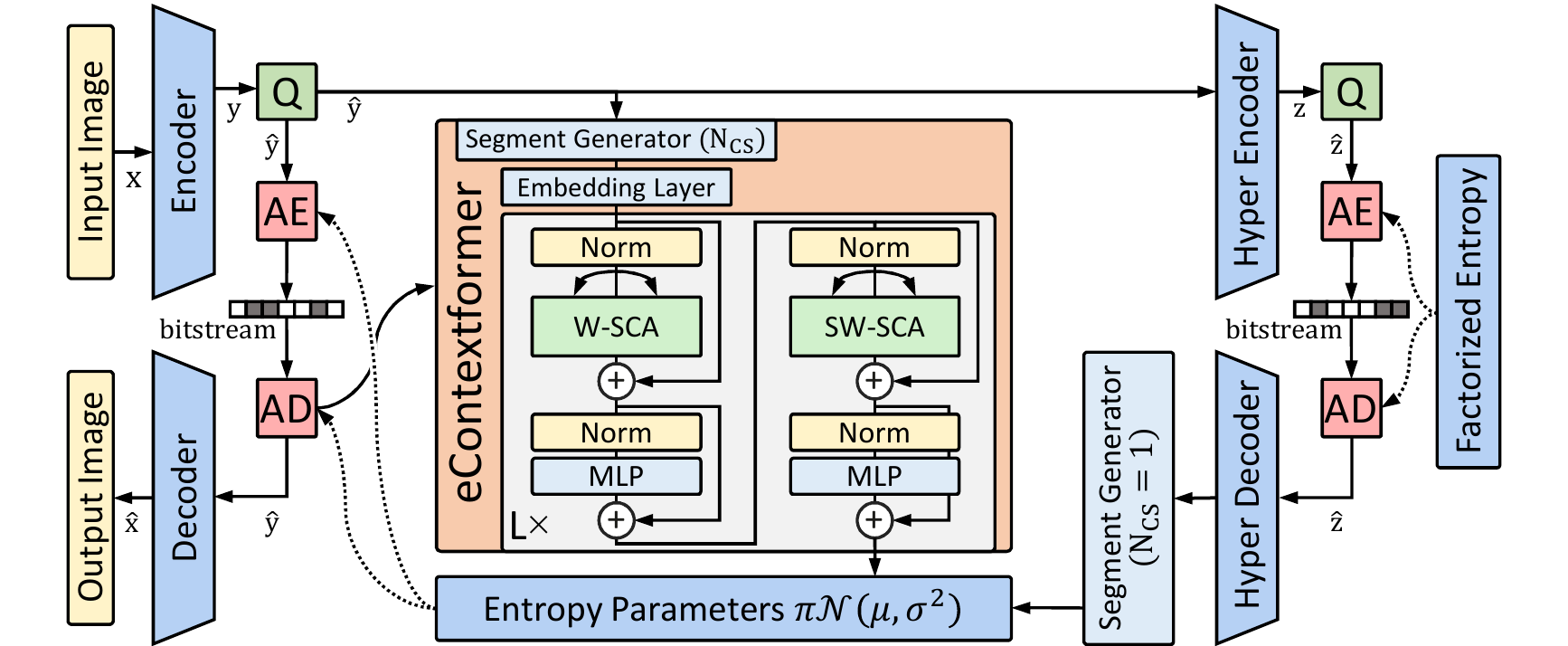}
\caption{Illustration of our compression framework utilizing with the eContextformer with window and shifted-window spatio-channel attention. The segment generator splits the latent into $N_{cs}$ channel segments for further processing. Following our previous work~\cite{koyuncu2022contextformer}, the output of hyperdecoder is not segmented but repeated along channel dimension to include more channel-wise local neighbors for the entropy modeling.}
\label{fig_our_model}
\end{figure*}

Our experiments show that the Contextformer~\cite{koyuncu2022contextformer} can support the checkered and patch-wise parallelization techniques discussed in \cref{sec_2a} but require a more efficient implementation for overlapping window attention. To achieve this, we replaced the ViT-based transformer in Contextformer with a Swin-based one and added spatio-channel attention. \cref{fig_our_model} describes the overall architecture of our compression framework. For context modeling, the segment generator rearranges the latent  $\bm{\hat{y}}\in\mathbb{R}^{\frac{H}{16} \times \frac{W}{16} \times C}$ into a spatio-channel segmented representation $\bm{\hat{s}}\in\mathbb{R}^{N_{cs} \times \frac{H}{16} \times \frac{W}{16} \times p_{cs}}$ with the number of channel segments $N_{cs}{=}\frac{C}{p_c}$. A linear layer converts $\bm{\hat{s}}$  to tensor embedding with the last dimension of $d_e$. The embedding representation passes through $L$ Swin transformer layers where window and shifted-window spatio-channel attention (W-SCA and SW-SCA) are applied in an alternating fashion. Window attention is computed on the intermediate outputs with $\mathbb{R}^{N_{w} \times (N_{cs}K^2) \times p_{cs}}$, where the first and second dimension represents the number of windows $N_{w}{=}\frac{H W}{256 K^2}$ processed in parallel and the sequential data within a window of $K{\times}K$, respectively. We masked each window according to the group coding order (see \cref{fig_ctx_models2}) to ensure coding causality, where each group uses all the previously coded groups for the context modeling. Following~\cite{liu2021swin}, we replaced absolute positional encodings with relative ones to introduce more efficient permutation variance. We employ multi-head attention with $h$ heads to yield better parallelization and build independent relations between different parts of the spatio-channel segments. More detailed description of our compression framework with the eContextformer is provided in \cref{tab:model_description}.

\begin{table}
    % \ra{1.3}
    \addtolength{\tabcolsep}{-4pt}
    
	\centering
	\caption{Improvements of proposed method over the Contexformer~\cite{koyuncu2022contextformer}}
	\label{tab_comp}
	\begin{tabular}{@{}lcc@{}}

		\toprule
		& Contextformer~\cite{koyuncu2022contextformer} & eContextformer  \\
		\midrule
		 Model architecture & ViT~\cite{dosovitskiy2020image} inspired & Swin~\cite{liu2021swin} inspired\\
		 \midrule
	     Attention mechanism &  & (expanding) \\
	      & spatio-channel & spatio-channel \\
	      & sliding window & shifted window \\
	      \midrule
	     Attention span &  &  \\
	     \quad spatial& $16{\times}16$ & $8{\times}8$ \\
	     \quad channel-wise& $4$ & $4$ \\
	      \midrule
	     Processing type & serial & parallel \\
	     	      &  & by fusing~\cite{koyuncu2021parallel,he2021checkerboard,minnen2020channel} \\
	     \quad \# autoregressive steps & $\frac{HW}{64}$ & 7\\
	     \midrule
	     Additional Optimizations &  &  \\

	      \quad Decoding runtime & WPP  & EGR, SFG and Caching \\
	       & (43\% reduction) & (64\% reduction) \\
	      \quad Complexity& \xmark  & EGR, SFG and Caching\\
	       & & (84\% reduction)\\
	      \midrule
	      Supports oRDO& \xmark & \cmark \\
		\bottomrule
	\end{tabular}
	   % \addtolength{\tabcolsep}{-4pt}
\end{table}
\cref{tab_comp} summarizes the improvements of eContextformer over the Contextformer~\cite{koyuncu2022contextformer}. The proposed (S)W-SCA mechanisms can be seen as an efficient fusion of the parallelization techniques~\cite{koyuncu2021parallel,he2021checkerboard,minnen2020channel} while not requiring additional CNNs and RNNs for aggregating channel-wise dependencies~\cite{he2022elic,minnen2020channel} and spatial dependencies~\cite{koyuncu2021parallel}. We further optimize the model architecture to lower model complexity even more to enable oRDO, which will be discussed in the following sections.

\subsection{Additional Complexity Optimizations}
\label{sec_optimization}
\begin{figure*}[!t]
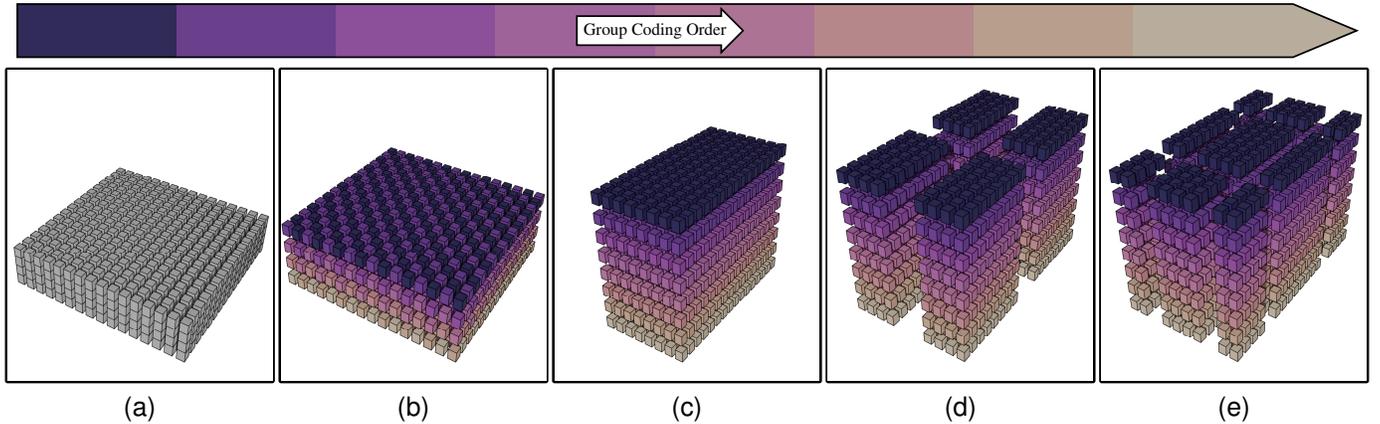

\centering
% \captionsetup[subfigure]{labelformat=empty}
\tcbset{rounded corners,colframe=black,left=0mm,right=0mm,top=0mm,bottom=0mm,boxsep=0mm,arc=0mm,boxrule=1pt}
% \subfloat[]{\tcbox{\input{figures/codingorder2.pgf}}}
\includegraphics[width=\textwidth]{figures/codingorder2.pdf}\vspace{-0.9\baselineskip}
\subfloat[]{\tcbox{\includegraphics[page=2,width=0.194\textwidth,trim={4.5cm 10.8cm 3.3cm 3.4cm},clip]{figures/model7.pdf}}}
\hfill
\subfloat[]{\tcbox{\includegraphics[page=3,width=0.194\textwidth,trim={4.6cm 10.8cm 3.2cm 3.4cm},clip]{figures/model7.pdf}}}
\hfill
\subfloat[]{\tcbox{\includegraphics[page=4,width=0.194\textwidth,trim={2.4cm 13.0cm 5.4cm 1.2cm},clip]{figures/model7.pdf}}}
\hfill
\subfloat[]{\tcbox{\includegraphics[page=5,width=0.194\textwidth,trim={2.4cm 11.15cm 5.4cm 3.05cm},clip]{figures/model7.pdf}}}
\hfill
\subfloat[]{\tcbox{\includegraphics[page=6,width=0.194\textwidth,trim={2.4cm 11.15cm 5.4cm 3.05cm},clip]{figures/model7.pdf}}}
\caption{Illustration of the optimized processing steps of eContextformer. From left to right, the latent tensor (a) is first split into channel segments (b) and reordered according to group coding order (c). Finally, the transformer layers with window and shifted-window spatio-channel attention (d-e) are applied on the reordered tensor, sequentially.}
\label{fig_optimization}
\end{figure*}

To generate the windows and shifted windows, the transformations $\mathbb{R}^{N_{cs} \times \frac{H}{16} \times \frac{W}{16} \times d_{e}} \leftrightarrow \mathbb{R}^{N_{w} \times (N_{cs}K^2) \times d_{e}}$ are applied to intermediate outputs in pre-post attention layers. Since eContextformer is autoregressive in spatial and channel dimensions, one can iterate over the channel segments with a checkered mask during decoding, i.e., by passing previous segments $\bm{\hat{s}}_{<i}$ through the context model to decode $\bm{\hat{s}}_{i}$ (with the channel segment index $i$). For instance, in the spatial-first-order (sfo) setting, two autoregressive steps are required per iteration, while the computation of half of the attention map is unnecessary for each first step. To remedy this, we rearranged the latent tensors into a form $\bm{\hat{r}}\in\mathbb{R}^{2N_{cs} \times \frac{H}{16} \times \frac{W}{32} \times p_{cs}}$ that is more suitable for efficient processing. We set the window size to $K{\times}\frac{K}{2}$ to preserve the spatial attention span, and applied the window and shifted attention as illustrated in \cref{fig_optimization}. We refer to the proposed rearrangement operation as Efficient coding Group Rearrangement (EGR). Furthermore, we code the first group $\bm{\hat{r}}_1$ only using the hyperprior instead of using a start token for context modeling of $\bm{\hat{r}}_1$. This reduced the total autoregressive steps for using the context model to $2N_{cs}-1$, i.e., a total complexity reduction of $13\%$ for $N_{cs}{=}4$. We refer to this optimization as Skipping First coding Group (SFG).

Note that transformers are sequence-to-sequence models and compute attention between every query and key-value pair. Let $\bm{Q}(1\le n)$, $\bm{K}(1\le n)$, and $\bm{V}(1\le n)$ be all queries, keys, and values, and $Attn(\bm{Q}(1\le n),\bm{K}(1\le n),\bm{V}(1\le n))$ be the attention computed up to coding step $n$. Since context modeling is an autoregressive task, during decoding, one can omit to compute the attention between the previously coded queries and key-value pairs and compute simply the attention $Attn(\bm{Q}(n),\bm{K}(1\le n),\bm{V}(1\le n))$ between the current query and the cached key-value pairs. For efficient implementation, we also adopted the key-value caching according to the white paper~\cite{scale2020} and the published work~\cite{yan2021fastseq}.

The combined implementation of the EGR and key-value caching reveals an emergent property of the proposed attention mechanism. At its foundation, the attention mechanism resembles the shifted window attention of~\cite{liu2021swin} for each coding step; ergo, $\bm{\hat{r}}_i$. However, the attention span expands with each coding step in spatial and channel dimensions. Therefore, our optimized attention mechanism can also be called as an expanding (S)W-SCA (see \cref{tab_comp}). Additionally, the caching efficiently reduces the complexity from $\mathcal{O}(N^2)$ to $\mathcal{O}(N)$ for the growing attention span.
%backup:Generation the shifted and non-shifted windows requires input to be in a \textit{planer} form while attention is calculated on a sequential data. Therefore, the transformations $\mathbb{R}^{N_{cs} \times H \times W \times d_{e}} \leftrightarrow \mathbb{R}^{N_{w} \times (N_{cs}K^2) \times d_{e}}$ are applied to intermediate outputs in pre-post attention layers. Since eContextformer is autoregressive in spatial and channel dimension, one can iterate over the channel segments during decoding, i.e. pass previous segments $\bm{\hat{y}_s}(<i)$ through the context model to decode (with the channel segment index $i$.

\section{Experiments}
\subsection{Experimental Setup}

\paragraph{Training} We configured eContextformer using the following parameters $\{N{=}192,\,M{=}192,\,N_{cs}{=}4,\,K{=}8,\allowbreak L{=}8,\,d_e{=}\frac{8M}{N_{cs}},\,d_{mlp}{=}4d_e,\,h{=}12\}$. Here, $N$, $M$, $N_{cs}$, and $d_e$ stand for the intermediate layer size of the encoder and decoder, the bottleneck size, the number of channel segments, and the embedding size. $L$ denotes the total number of transformer layers, where half utilize W-SCA and the other half has SW-SCA. Those parameters were selected to be identical to the ones used in Contextformer~\cite{koyuncu2022contextformer}, in order to ensure a fair comparison. Since our initial experiments showed deteriorating performance for the $cfo$ coding method for window attention, we continued with the $sfo$ version. $K$ defines the size of the attention span in the spatial dimension, which results in $K{\times}\frac{K}{2}$ window for the proposed optimization algorithms (see \cref{sec_optimization}). Following~\cite{koyuncu2022contextformer,begaint2020compressai}, we trained eContextformer for 120 epochs (${\sim}$1.2M iterations) on $256{\times}256$ random image crops with a batch size of 16 from the Vimeo-90K dataset~\cite{xue2019video}, and used ADAM optimizer~\cite{kingma2014adam} with the initial learning rate of $10^{-4}$. In order to cover a range of bitrates, we trained various models with $\lambda\in \{0.002,\,0.004,\,0.007,\,0,014,\,0.026,\,0.034,\,0.058\}$ with mean-squared-error (MSE) as the distortion metric $\mathbf{D}(\cdot)$. To evaluate the perceptual quality of our models,  we finetuned them with MS-SSIM \cite{wang2003multiscale} as the distortion metric for ${\sim}$500K iterations. For high bitrate models ($\lambda_{5,6,7}$), we increased the bottleneck size to 312 according to the common practice \cite{cheng2020learned,minnen2018joint,minnen2020channel}. Empirically, we observed better results with a lower number of heads for the highest rate model, so we used 6 heads for this model. Additionally, we investigated the effect of the training dataset and crop size by finetuning the models with $256{\times}256$ and $384{\times}384$ image crops from the COCO~2017 dataset~\cite{lin2014microsoft} for ${\sim}$600K iterations.
\begin{figure*}[!t]
\centering
\subfloat[\label{fig_exp1_a}]{\includegraphics[width=0.5\linewidth]{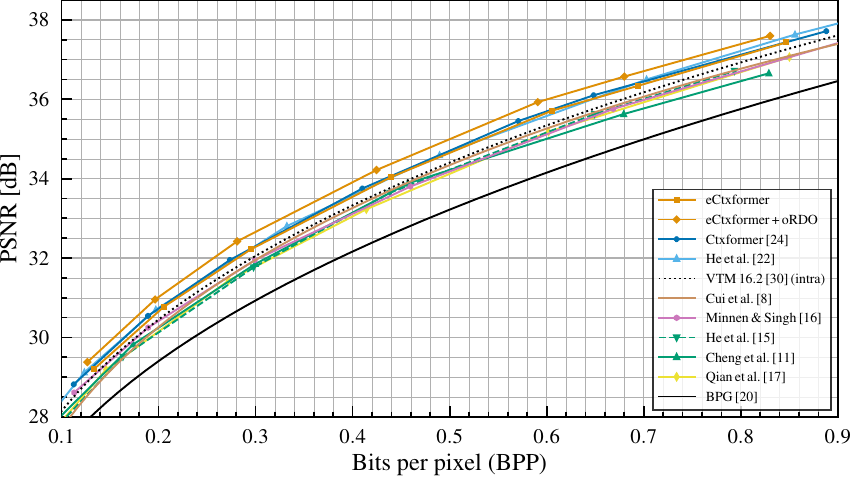}}\hfill
\subfloat[\label{fig_exp1_b}]{\includegraphics[width=0.5\linewidth]{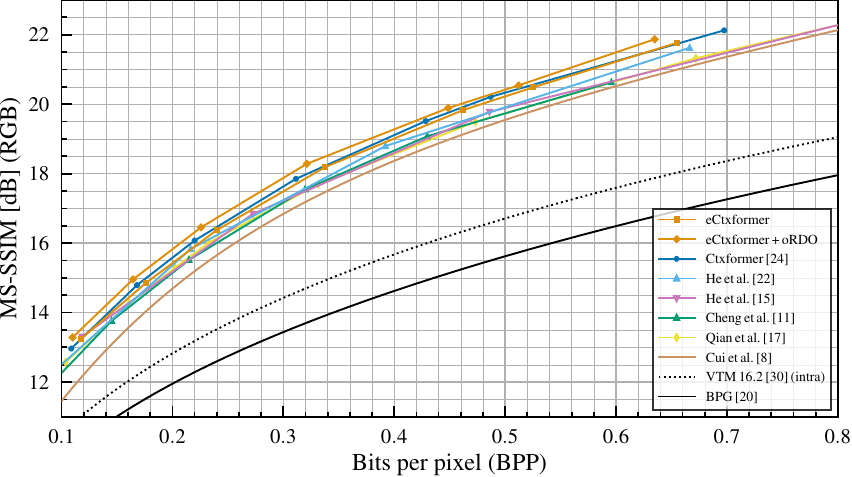}}
\caption{The rate-distortion performance in terms of (a) PSNR and (b) MS-SSIM on Kodak dataset \cite{franzen1999kodak} showing the performance of our model compared to various learning-based and classical codecs. We also include the performance of our model combined with oRDO.}
\label{fig_exp1}
\end{figure*}

\begin{figure*}[!t]
\centering
\subfloat[]{\includegraphics[width=0.5\linewidth]{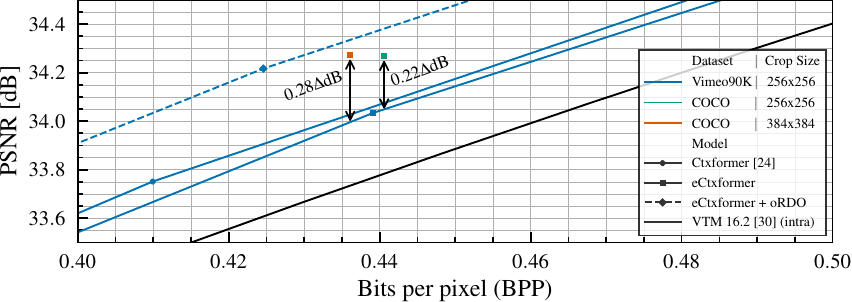}}\hfill
\subfloat[]{\includegraphics[width=0.5\linewidth]{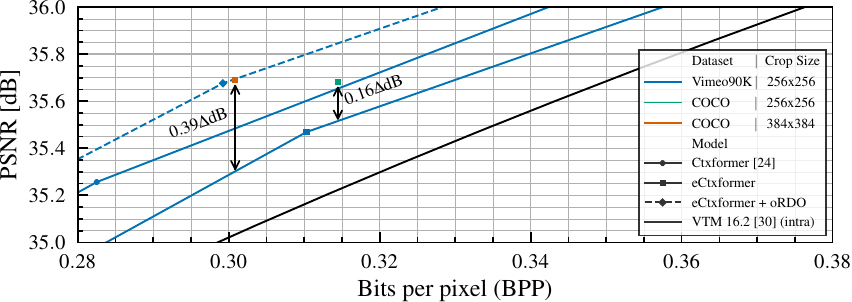}}
\caption{Experimental study of the effects of different training datasets and crop sizes on the performance, showing the rate-distortion performance on (a) Kodak \cite{franzen1999kodak} and (b) Tecnick \cite{asuni2014testimages} datasets}
\label{fig_abl2}
\end{figure*}

\paragraph{Evaluation} We analyzed the performance of the eContextformer on the Kodak image dataset \cite{franzen1999kodak}, CLIC2020 \cite{CLIC2020} (Professional and Mobile) test dataset, and Tecnick \cite{asuni2014testimages} dataset. We compared its performance to various serial and parallel context models: the 2D context models (Minnen et al.~\cite{minnen2018joint}, Cheng et al.~\cite{cheng2020learned} and Fu et al.~\cite{fu2023asymmetric}), the multi-scale 2D context model (Cui et al.~\cite{cui2021asymmetric}), the 3D context models (Chen et al.~\cite{liu2019non} and Tang et al.~\cite{tang2022joint}), the context model with a simplified attention (Guo et al.~\cite{guo2021causal}), the transformer-based context model with spatio-channel attention (Contextformer~\cite{koyuncu2022contextformer}), the channel-wise autoregressive context model (Minnen\&Singh~\cite{minnen2020channel}), the checkerboard context model (He et al.~\cite{he2021checkerboard}), transformer-based checkerboard context model (Qian et al. \cite{qian2021entroformer}), the channel-wise and checkerboard context model (He et al.~\cite{he2022elic}), the transformer-based channel-wise and checkerboard context model (Liu et al.~\cite{liu2023learned}). We tested the \textit{small} (S) and \textit{large} (L) models of Liu et al.~\cite{liu2023learned}, which differ in the number of intermediate layer channels. Additionally, we  experimented with the low-complexity asymmetric frameworks of Wang et al.~\cite{wang2023evc}, Fu et al.~\cite{fu2023asymmetric} and Yang\&Mandt~\cite{yang2023computationally}. We used the model (LL) from~\cite{wang2023evc} with a \textit{large} synthesis and \textit{large} analysis transforms since it has a higher rate-distortion performance. We also included the results of the framework from~\cite{yang2023computationally} using an oRDO called Stochastic Gumbel Annealing (SGA)~\cite{yang2020improving} that applies iterative optimization for 3000 steps. If the source was present, we executed the inference algorithms of all those methods; otherwise, we obtained the results from relevant publications. We also used classical image coding frameworks such as BPG~\cite{bellard2015bpg} and VTM~16.2~\cite{jvet2019versatile} (intra) for comparison. 

Moreover, we measured the number of parameters of entropy models with the summary functions of PyTorch~\cite{pytorch} or Tensorflow~\cite{tensorflow} (depending on the published implementation). By following the recent standardization activity JPEG-AI~\cite{ascenso2023jpeg}, we computed the model complexity in kMAC/px with the ptflops package~\cite{ptflops}. In case of missing hooks for attention calculation, we integrated them with the help of the official code repository of the Swin transformer~\cite{liu2021swin}. For the runtime measurements, we used DeepSpeed~\cite{rasley2020deepspeed} and excluded arithmetic coding time for a fair comparison since each framework uses a different implementation of the arithmetic codec. All the tests, including the runtime measurements, are done on a machine with a single NVIDIA Titan RTX and Intel Core~i9-10980XE. We used PyTorch~1.10.2~\cite{pytorch} with CUDA Toolkit~11.4~\cite{cuda}. 

\subsection{Model Performance}
\cref{fig_exp1} shows the rate-distortion performance of the eContextformer trained on the Vimeo-90K with $256{\times}256$ image crops. In terms of PSNR, our model qualitatively
outperforms the VTM~16.2~\cite{jvet2019versatile} for all rate points under test and achieves 5.3\% bitrate savings compared to it (see \cref{fig_exp1_a}). Our compression framework shares the same analysis, synthesis transforms, and hyperprior with the model with multi-scale 2D context model~\cite{cui2021asymmetric} and Contextformer~\cite{koyuncu2022contextformer}. Compared to~\cite{cui2021asymmetric}, our model saves 8.5\% more bits, while it provides 1.7\% lower performance than Contextformer due to the parallelization of context modeling. The eContextformer achieves competitive performance to parallelized models employing channel-wise autoregression with an increased number of model parameters, such as Minnen\&Singh~\cite{minnen2020channel} and He et al.~\cite{he2022elic}. When comparing with VTM~16.2~\cite{jvet2019versatile}, the former model gives 1.6\% loss in BD-Rate performance, and the latter gives 6.7\% gain.

In \cref{fig_exp1_b}, we also evaluated the perceptually optimized models compared to the prior art. In terms of MS-SSIM \cite{wang2003multiscale}, the eContextformer saves on average 48.3\% bitrate compared to VTM~16.2~\cite{jvet2019versatile}, which is performance-wise 0.8\% better than He et al.~\cite{he2022elic} and 1.1\% worse than the Contextformer~\cite{koyuncu2022contextformer}.

\begin{figure*}[!t]
\centering
\subfloat[]{\includegraphics[width=0.32\linewidth]{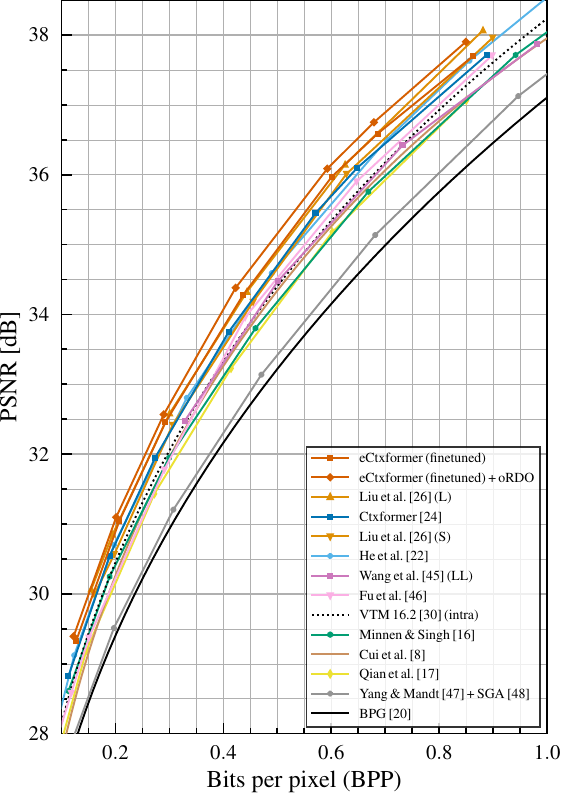}}\hfill
\subfloat[]{\includegraphics[width=0.32\linewidth]{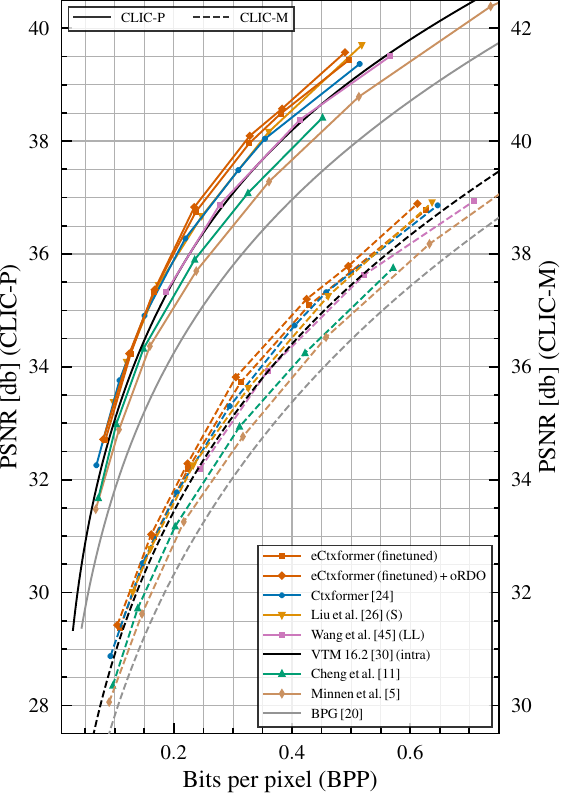}}\hfill
\subfloat[]{\includegraphics[width=0.32\linewidth]{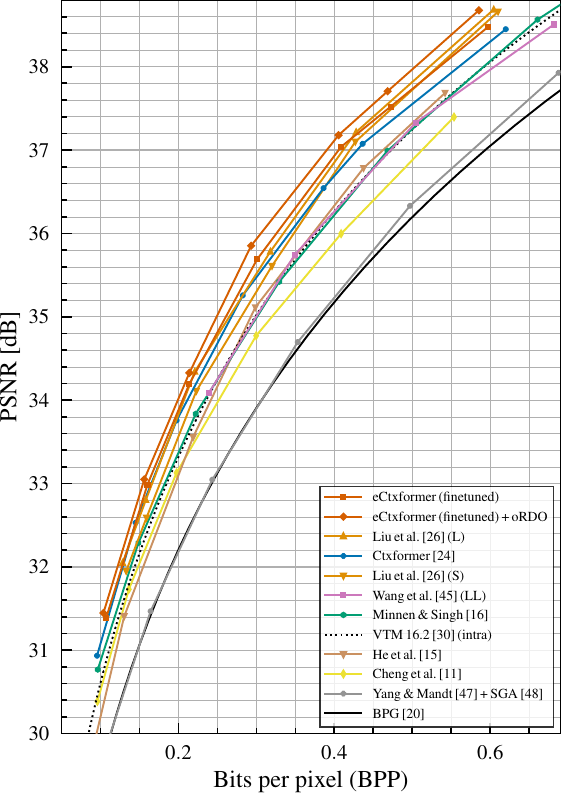}}
\caption{The rate-distortion performance in terms of PSNR on (a) Kodak \cite{franzen1999kodak}, (b) CLIC2020 \cite{CLIC2020} (Professional and Mobile), and (c) Tecnick \cite{asuni2014testimages} datasets, showing the performance of our model w/ and w/o finetuning compared to various learning-based and classical codecs. We also include the performance of our models combined with oRDO.}
\label{fig_exp2}
\end{figure*}
\subsection{Effect of training with larger image crops}
Although eContextformer yields a performance close to Contextformer on the Kodak dataset \cite{franzen1999kodak}, it underperforms on larger resolution  test datasets such as Tecnick~\cite{asuni2014testimages} (see \cref{fig_abl2}. The Vimeo-90K dataset~\cite{xue2019video} we initially used for our training has a resolution of $448{\times}256$. In order to avoid a bias towards horizontal images, we used $256{\times}256$ image crops, ergo  $16{\times}16$ latent variable, for training similar to the state-of-the-art~\cite{begaint2020compressai}. However, the attention window size of $K{=}8$ combined with low latent resolution limits learning an efficient context model. In order to achieve high rate-distortion gain, the recent studies \cite{koyuncu2021parallel,minnen2020channel,he2022elic,qian2021entroformer,fu2023asymmetric} experimented with higher resolution datasets, such as ImageNet~\cite{deng2009imagenet}, COCO~2017~\cite{lin2014microsoft}, and DIV2K~\cite{agustsson2017ntire}, and larger crop sizes up to $512{\times}512$. Following those studies, we finetuned eContextformer with $256{\times}256$ and $384{\times}384$ image crops from COCO~2017~\cite{lin2014microsoft}. The models finetuned with different crop sizes achieve similar performance on the Kodak dataset \cite{franzen1999kodak}, which has about 0.4M pixels. On the contrary, larger crop sizes help the models to reach more than two times in performance on the Tecnick dataset~\cite{asuni2014testimages}, which has ${\sim}1.5$M pixels per image.

\cref{fig_exp2} and \cref{tab_perf} show the rate-distortion performance of the finetuned eContextformer on Kodak \cite{franzen1999kodak}, CLIC2020 \cite{CLIC2020} (Professional and Mobile), and Tecnick \cite{asuni2014testimages} datasets. Our finetuned models reach a compression performance on par with the state-of-the-art LIC frameworks, providing over 5\% bitrate saving over the initial training and achieving average savings of 10.9\%, 12.3\%, 6.9\%, and 13.2\% over VTM 16.2~\cite{jvet2019versatile} on those datasets, respectively. 

% \addtolength{\tabcolsep}{4pt}    
\begin{table}
    % \ra{1.3}
\newcolumntype{x}[1]{>{\centering\arraybackslash\hspace{0pt}}p{#1}}

    \addtolength{\tabcolsep}{-2pt}

	\centering
	\newcommand{\noentrya}{\multicolumn{2}{c}{\phantom{11}\textendash\phantom{$^b$}}}
	\newcommand{\noentryb}{\multicolumn{2}{c}{\phantom{11}\textendash$^b$}}
	\newcommand{\noentryc}{\phantom{11}\textendash$^b$}

\begin{threeparttable}
	\caption{Performance of our model compared to various various learning-based codecs for various datasets}
	\label{tab_perf}
	\begin{tabular}{@{}lcccc@{}}
		\toprule
		& \multicolumn{4}{c}{BD-Rate [\%]$^a$ over VTM 16.2~\cite{jvet2019versatile} (intra)} \\
	    \cmidrule(lr){2-5}
	    & & \multicolumn{2}{c}{CLIC2020~\cite{CLIC2020}}&\\
	    \cmidrule(lr){3-4}
		Method & \mc{Kodak~\cite{franzen1999kodak}} & -P & -M & \mc{Tecnick~\cite{asuni2014testimages}} \\
		\midrule
		eContextformer & $-10.9$ & $-12.3$ & $\phantom{1}\!\!-6.9$ & $-13.2$ \\
		\quad + oRDO         & $-14.7$ & $-14.1$ & $-10.4$ & $-17.0$ \\
		Contextformer~\cite{koyuncu2022contextformer} & $\phantom{1}\!\!-6.9$& $\phantom{1}\!\!-8.8$& $\phantom{1}\!\!-5.8$& $-10.5$ \\
		\midrule
		Liu et al.~\cite{liu2023learned} (L)              & $-11.0$& \noentrya& $-12.2$\\
		He et al.~\cite{he2022elic}                   & $\phantom{1}\!\!-6.7$& \noentrya & \textendash\\
		Liu et al.~\cite{liu2023learned} (S)              & $\phantom{1}\!\!-6.2$& $-9.8$ &$\phantom{1}\!\!-4.2$& $\phantom{1}\!\!-8.0$\\
		Guo et al.~\cite{guo2021causal}               & $\phantom{1}\!\!-3.7$ & \noentrya & \textendash\\
		Wang et al.~\cite{wang2023evc} (LL)               & $\phantom{1}\!\!-0.6$ & $\phantom{1}\!\!-1.6$ & $\phantom{-1}0.4$ & $\phantom{1}\!\!-0.7$ \\
		Fu et al.~\cite{fu2023asymmetric}             & $\phantom{1}\!\!-0.3$ & \noentrya & \textendash\\
		Minnen\&Singh \cite{minnen2020channel}        & $\phantom{-1}1.9$&  \noentrya & $\phantom{1}\!\!-2.2$ \\
		Cui et al.~\cite{cui2021asymmetric}           & $\phantom{-1}3.2$&  \noentrya & \textendash \\
		Cheng et al.~\cite{cheng2020learned}       & $\phantom{-1}4.2$&  $\phantom{-1}5.9$& $\phantom{-1}9.1$& $\phantom{-1}4.8$ \\
		He et al.~\cite{he2021checkerboard}           & $\phantom{-1}4.5$&  \noentrya & $\phantom{-1}1.5$ \\
		Qian et al.~\cite{qian2021entroformer}        & $\phantom{-1}4.8$&  \noentryb & \noentryc \\
		Tang et al.~\cite{tang2022joint}        & $\phantom{-1}5.7$&  \noentrya & $\phantom{-1}3.5$ \\
		Minnen et al.~\cite{minnen2018joint}          & $\phantom{-}14.7$& $\phantom{-}11.6$ &$\phantom{-}14.1$& $\phantom{-}15.4$ \\
		Yang\&Mandt~\cite{yang2023computationally}    & $\phantom{-}36.5$ & \noentrya& $\phantom{-}45.7$ \\
		\quad + SGA~\cite{yang2020improving}          & $\phantom{-}22.4$ & \noentrya&$\phantom{-}26.9$ \\
		BPG~\cite{bellard2015bpg}                     & $\phantom{-}30.3$& $\phantom{-}40.3$&$\phantom{-}30.9$& $\phantom{-}29.5$ \\
		\midrule
		VTM 16.2~\cite{jvet2019versatile} (intra)                     & $\phantom{-0}0.0$& $\phantom{-0}0.0$&$\phantom{-0}0.0$& $\phantom{-0}0.0$ \\
		\bottomrule
	\end{tabular}
\begin{tablenotes}
\item[$^a$] Lower means higher bitrate savings
\item[$^b$] Can not be computed due to out of memory (OOM)
\end{tablenotes}
\end{threeparttable}
\end{table}

\subsection{Model and Runtime Complexity}
\cref{tab_opt} shows the complexity of the eContextformer w.r.t. different optimization proposed in \cref{sec_optimization}. During encoding and decoding, each of the EGR and SFG methods decreases the complexity of context modeling by 10-13\%, whereas the combination of them with the caching of key-value pairs provides an 84\% complexity reduction in total. We also compared the efficiency of caching to the single pass, where the whole latent variable $\bm{\hat{y}}$ is processed on the encoder side at once with the given causality mask. The caching is also more efficient than the single pass since only the required parts of the attention map are calculated. 
In~\cref{tab_modelsize}, we compared the number of model parameters and the complexity of the entropy model (including $g_{c}$, $h_{a}$, $h_{s}$, and $g_{e}$) of our model and some of the prior arts. Compared to the other transformer-based methods, the proposed window attention requires less computation than the sliding window attention of Contextformer and global attention of~\cite{qian2021entroformer}. For instance, with the proposed optimizations enabled, eContextformer has 145x lower complexity than the Contextformer for a similar number of model parameters. The spatio-channel window attention can efficiently aggregate information of channel segments without concatenation. Therefore, our model requires a smaller and shallower entropy parameter network compared to~\cite{cui2021asymmetric}, and a significantly lower total number of parameters in the entropy model compared to channel-wise autoregressive models~\cite{minnen2020channel}.

\begin{table}
    % \ra{1.3}
    % \addtolength{\tabcolsep}{8pt}
    
	\centering
	\caption{Ablation study of the proposed optimization methods applied to eContextformer}
	\label{tab_opt}
 \begin{tabular}{@{}cccc|c@{}}

		\toprule
		EGR & SFG  & Single Pass & Caching & kMAC/px  \\
		\midrule
		 \xmark& \xmark& \xmark& \xmark& 1200  \\
	     \cmark & \xmark& \xmark& \xmark& 1075 \\
	     \cmark & \cmark  & \xmark& \xmark& \phantom{0}829 \\
	     \cmark & \cmark  & \cmark & \xmark& \phantom{0}212 \\
	     \cmark & \cmark  & \xmark& \cmark & \phantom{0}204 \\
		\bottomrule
	\end{tabular}
	   % \addtolength{\tabcolsep}{-4pt}

\end{table}
% \addtolength{\tabcolsep}{4pt}    
\begin{table}
    % \ra{1.3}

    \addtolength{\tabcolsep}{-3pt}

	\centering
\begin{threeparttable}
	\caption{Number of parameters and entropy model complexity of our model compared to various various learning-based codecs}
	\label{tab_modelsize}
	\begin{tabular}{@{}lccc@{}}
		\toprule
		& \multicolumn{2}{c}{\# Parameters $[\text{M}]$} & \multicolumn{1}{c}{Complexity [kMAC/px]$^a$} \\
	    \cmidrule(lr){2-3}         \cmidrule(lr){4-4}
		& \mc{Analysis\&} & \mc{Entropy}& \mc{Entropy} \\
		Method& \mc{Synthesis} & \mc{Model}& \mc{Model} \\
		\midrule 
		eContextformer (opt.) $^b$ & 17.5 & \phantom{1}26.9 & 253 \\
		Contextformer~\cite{koyuncu2022contextformer} & 17.5 & \phantom{1}20.1& {$36\cdot10^3$} \\
		\midrule
		Liu et al.~\cite{liu2023learned} (L) $^{c}$& 28.5 & \phantom{0}47.4 & 125 \\
		Liu et al.~\cite{liu2023learned} (S) & \phantom{1}8.8 & \phantom{0}36.2 & 116 \\
		He et al.~\cite{he2022elic} & 14.7 & \phantom{0}44.8 & 266 \\
		Cui et al.~\cite{cui2021asymmetric} & 17.5 &\phantom{1}21.8 & \phantom{0}56 \\
		Qian et al.~\cite{qian2021entroformer} $^c$& \phantom{1}7.6 &\phantom{1}29.0 & 868\\
		Minnen\&Singh \cite{minnen2020channel}  & \phantom{1}8.4 &113.5& 426\\
		Yang\&Mandt~\cite{yang2023computationally} & \phantom{0}8.6 & \phantom{0}15.2 & \phantom{0}22 \\
		Wang et al~\cite{wang2023evc} (LL) & \phantom{0}6.6 & \phantom{0}10.8 & \phantom{0}37 \\
		Minnen et al.~\cite{minnen2018joint}& \phantom{1}5.6 &\phantom{11}8.3& \phantom{0}24 \\	
		\bottomrule
	\end{tabular}
\begin{tablenotes}
\footnotesize{
\item[$a$] The complexity is measured on 4K images with 3840x2160 resolution.
\item[$b$] The combination of the proposed EGR, SFG and Caching optimizations are applied.
\item[$c$] The complexity is approximated by using images with 1024x1024 resolution due to out of memory (OOM)

}\end{tablenotes}
\end{threeparttable}

\end{table}

\cref{tab_complexity} presents the encoding and decoding runtime complexity of our model, some of the prior learning-based models and VTM~16.2~\cite{jvet2019versatile}. The proposed optimizations speed up the encoding and decoding up to 3x, proving a 210x improvement over the optimized version of the Contextformer~\cite{koyuncu2022contextformer}. Furthermore, we observed that coding time scales better for the optimized eContextformer compared to the one without the proposed optimizations. The 4K images have 21x more pixels than the Kodak images, while the relative encoding and decoding time of the optimized models for those images increase only 14x w.r.t. the ones on the Kodak dataset. Moreover, our optimized model provides competitive runtime performance to the Minnen\&Singh~\cite{minnen2020channel} and both small and large models of Liu et al.~\cite{liu2023learned} (S/L).

We also observed that the memory usage in~\cite{qian2021entroformer} increases significantly with the image resolution due to the expensive global attention mechanism in their context model. Similarly, in~\cite{liu2023learned}, the residual transformer-based layers in synthesis and analysis transforms, and the context model combined with the channel-wise grouping put a heavy load on memory. Therefore, we could not test Liu et al.~\cite{liu2023learned} (L) and Qian et al.~\cite{qian2021entroformer} on 4K images on our GPU with 24 GB memory.
\begin{table}
\begin{threeparttable}
    % \addtolength{\tabcolsep}{6pt}    

	\centering
	\caption{Encoding and decoding time of our model compared to various learning-based and classical codecs}
	\label{tab_complexity}
	\begin{tabular}{@{}lcccc@{}}
		\toprule
		& \multicolumn{2}{c}{Enc. Time [s]}& \multicolumn{2}{c}{Dec. Time [s]} \\
		\cmidrule(lr){2-3} 		\cmidrule(lr){4-5}
		Method & Kodak & 4K$^a$&Kodak &4K$^a$ \\
		\midrule
		eContextformer (opt.) $^b$                                 & \phantom{00}0.16  & \phantom{000}2.24\phantom{$^d$} & \phantom{00}0.17 & \phantom{000}2.41\phantom{$^b$} \\
		eContextformer                                             & \phantom{00}0.33  & \phantom{000}6.55\phantom{$^d$} & \phantom{00}0.35 & \phantom{000}6.75\phantom{$^b$}  \\
		Contextformer~\cite{koyuncu2022contextformer}              & \phantom{0}56\phantom{.00}  & 1240\phantom{.00}\phantom{$^d$} & \phantom{0}62\phantom{.00} & 1440\phantom{.00}\phantom{$^b$} \\
		Contextformer (opt.)~\cite{koyuncu2022contextformer} $^c$  & \phantom{00}8.72  & \phantom{0}120\phantom{.00}\phantom{$^d$} & \phantom{0}44\phantom{.00} & \phantom{0}820\phantom{.00}\phantom{$^b$} \\
		\midrule
		Yang\&Mandt~\cite{yang2023computationally}                  &\phantom{00}0.09  & \phantom{00}0.81\phantom{$^d$} & \phantom{00}0.02 & \phantom{00}0.43\phantom{$^d$} \\
	    Wang et al.~\cite{wang2023evc} (LL)                             &\phantom{00}0.04  & \phantom{00}0.56\phantom{$^d$} & \phantom{00}0.04 & \phantom{00}0.57\phantom{$^d$} \\
		Minnen\&Singh~\cite{minnen2020channel}                     &\phantom{00}0.16   & \phantom{00}1.99\phantom{$^d$} & \phantom{00}0.12 & \phantom{00}1.37\phantom{$^b$} \\
	    Liu et al.~\cite{liu2023learned} (S)                          &\phantom{00}0.18  & \phantom{00}1.74\phantom{$^d$} & \phantom{00}0.17 &  \phantom{00}1.47\phantom{$^d$} \\
	    Liu et al.~\cite{liu2023learned} (L)                      &\phantom{00}0.25  & \textendash$^d$ & \phantom{00}0.23 & \textendash$^d$ \\
		Qian et al.~\cite{qian2021entroformer}                     & \phantom{00}1.57  & \textendash$^d$ & \phantom{00}2.59 & \textendash$^d$ \\
		Cheng et al.~\cite{cheng2020learned}                       & \phantom{00}2.22	& \phantom{00}54\phantom{.00}\phantom{$^b$} & \phantom{00}5.81 &\phantom{0}140\phantom{.00}\phantom{$^b$} \\
		Chen et al.~\cite{liu2019non}                              & \phantom{00}4.11	& \phantom{00}28\phantom{.00}\phantom{$^b$} & 316\phantom{.00}			 &7486\phantom{.00}\phantom{$^b$} \\
		VTM 16.2 \cite{jvet2019versatile} (intra)                         & 420\phantom{.00} 	& \phantom{0}950\phantom{.00}\phantom{$^b$}& \phantom{00}0.84			 &\phantom{000}2.53\phantom{$^b$} \\
		\bottomrule
	\end{tabular}
% 	\addtolength{\tabcolsep}{-6pt}
\begin{tablenotes}
\footnotesize{
\item[$a$] The time is measured on 4K images with 3840x2160 resolution.
\item[$b$] The combination of the proposed EGR, SFG and Caching optimizations are applied. Less than 0.1\% change in BD-Rate performance is observed due to the complexity optimizations.
\item[$c$] Complexity optimization techniques proposed in~\cite{koyuncu2022contextformer} are applied.
\item[$d$] Can not be computed due to out of memory (OOM)
}
\end{tablenotes}
\end{threeparttable}

\end{table}

\subsection{Online Rate-Distortion Optimization}
\begin{figure*}[!t]
\centering
\subfloat[\label{fig_abl5}]{\includegraphics[width=0.307\linewidth]{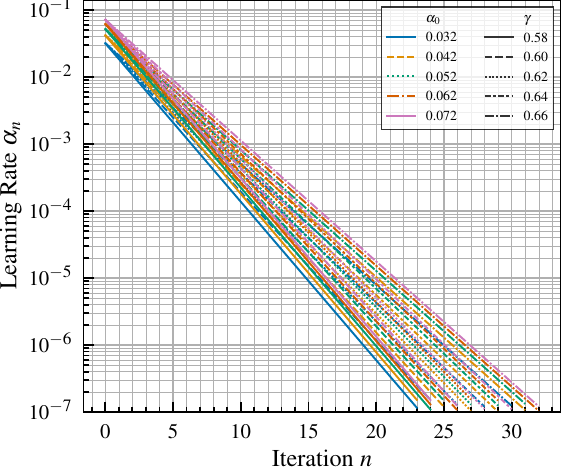}}\hfill
\subfloat[\label{fig_abl6}]{\includegraphics[width=0.33\linewidth]{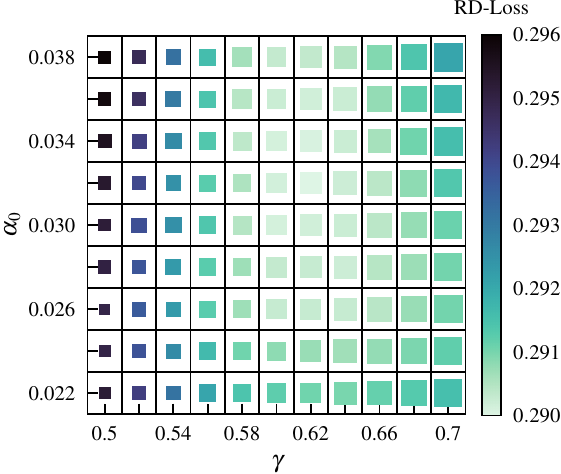}}\hfill
\subfloat[\label{fig_abl7}]{\includegraphics[width=0.33\linewidth]{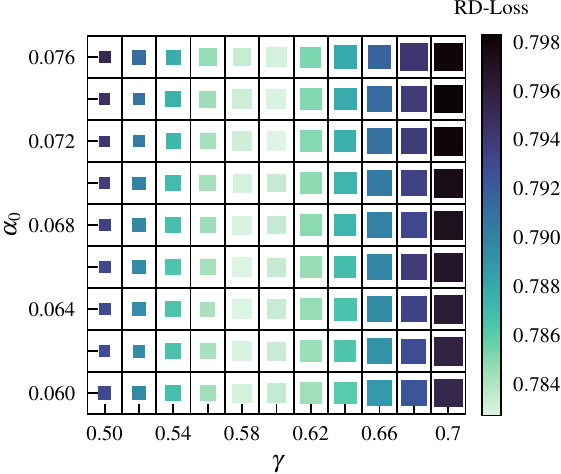}}
\caption{Illustration of (a) the learning rate decay used for the oRDO w.r.t. optimization iteration for different initial learning rate $\alpha_0$ and decay rate $\gamma$. The results of the TPE for our model (b) with $\lambda_1$ and (c) with $\lambda_4$ for different combinations of $(\alpha_{0},\gamma)$. The size of each square symbolizes the required number of oRDO iteration steps.}
\end{figure*}

Since using eContextformer in our framework results in a significantly lower encoder complexity, we could afford to incorporate an oRDO technique, which further improves the compression efficiency. We took the oRDO algorithm from~\cite{campos2019content}, replaced the noisy quantization with the straight-through estimator described in~\cite{bengio2013estimating}, and used an exponential decay for scheduling the learning rate, which results in a simplified oRDO with faster convergence. The learning rate at $n$-th iteration can be defined in closed-form as:
\begin{equation}
	\alpha_{n} = \alpha_{0}\gamma^n,
	\label{eq:rdo}
\end{equation}
where $\alpha_{0}$ and $\gamma$ are the initial learning rate and the decay rate, respectively. \cref{fig_abl5} illustrates the immediate learning rate $\alpha_{n}$ w.r.t. oRDO steps for different combinations of $(\alpha_{0},\gamma)$.

We obtained the optimal parameters by a search using Tree-Structured Parzen Estimator (TPE)~\cite{bergstra2011algorithms} with the objective function (\ref{eq_2}) and the constraints of $\alpha_{n}{>}10^{-7}$, $\alpha_{0}\in[0.02, 0.08]$, and $\gamma\in[0.5, 0.7]$. To omit over-fitting, we used 20 full-resolution images randomly sampled from the COCO~2017 dataset~\cite{lin2014microsoft} for the search. \cref{fig_abl6,fig_abl7} show the results of TPE~\cite{bergstra2011algorithms} for models trained with $\lambda_1$ and $\lambda_4$, respectively. We observed that the higher bitrate models ($\lambda_{>3}$) generally perform better with a higher initial learning rate compared to the ones trained for lower bitrates ($\lambda_{<4}$). This suggests that the optimization of less sparse latent space requires a larger step size at each iteration. We set $\alpha_{0}{=}0.062$ and $\gamma{=}0.72$ for $\lambda_{<4}$ and $\alpha_{0}{=}0.062$ and $\gamma{=}0.72$ for $\lambda_{>3}$, which results in 26 iteration steps for all models.

\cref{fig_exp1,fig_exp2} show the rate-distortion performance of the eContextformer with oRDO using the optimal parameter setting. Compared to VTM 16.2~\cite{jvet2019versatile}, the oRDO increases the rate-distortion performance of the finetuned models up to 4\%, providing 14.7\%, 14.1\%, 10.4\%, and 17.0\% average bitrate savings on Kodak \cite{franzen1999kodak}, CLIC2020 \cite{CLIC2020} (Professional and Mobile), and Tecnick \cite{asuni2014testimages} datasets, respectively. The encoding runtime complexity is proportional to the number of optimization steps used. For the selected number of steps, the total encoding runtime complexity of our model is still lower than the VTM 16.2~\cite{jvet2019versatile}. Furthermore, we observed that the oRDO increases the performance of our initial models (without the finetuning) by up to 7\%, which indicates those models are trained sub-optimally.

\subsection{Ablation Study}
\cref{tab_abl} summarizes the ablation study we conducted with Contextformer~\cite{koyuncu2022contextformer} and eContextformer on the Kodak dataset~\cite{franzen1999kodak}. Notably, the proposed architecture shares the same synthesis and analysis transforms with~\cite{koyuncu2022contextformer} and~\cite{cui2021asymmetric} and differs in the entropy modeling part. We observed that increasing the number of mixtures $k_m$ of GMM~\cite{cheng2020learned} provides 1.6-1.9\% performance gain for the Contextformer~\cite{koyuncu2022contextformer} and eContextformer, while the context model with spatio-channel attention increases the performance by more than 6\%. Compared to the straight-forward parallelization (pContextformer in \cref{sec_2}), the combination of proposed (S)W-SCA and complexity optimization techniques allows eContextformer to reach a better performance-complexity trade-off. The finetuning with larger resolution images is helpful for both eContextformer and Contextformer~\cite{koyuncu2022contextformer}. However, we observed that the sliding window attention is less stable while training with a higher resolution images, which might explain the lower performance gain for the Contextformer~\cite{koyuncu2022contextformer}. Moreover, finetuning reduces the discrepancy between the training and testing on different resolution images. Therefore, the performance contribution of oRDO to the finetuned model is lower than the one without the finetuning. 

Notably, the oRDO does not impact the entropy model complexity of the decoder since it is only applied during encoding. Each oRDO iteration requires one forward and one backward pass through the entropy model, where the backward pass approximately has two times more computations~\cite{Kaplan2019NotesOC,rasley2020deepspeed}. Therefore, the encoder-side entropy model complexity with $n$ oRDO iterations can be estimated as $3n$ times the encoder-side complexity without the oRDO. Moreover, for all models, the decoder-side entropy model complexity is slightly lower (${\sim}9$ kMAC/px) since the hyperprior's analysis transform ($h_a$) is not used during decoding.
\begin{table}
    % \ra{1.3}
    \addtolength{\tabcolsep}{-4.5pt}
    
	\centering
	\caption{Ablation study for eContextformer on Kodak dataset~\cite{franzen1999kodak}}
	\label{tab_abl}
 \begin{threeparttable}

	\begin{tabular}{@{}lcccc|cc|c@{}}

		\toprule
		Context  &          & GMM     & Fine- &      & \multicolumn{2}{c|}{kMAC/px$^b$} & BD-Rate  \\
		  Model & Type$^a$ & $k_m$  & tuning & oRDO &  Enc.& Dec.$\phantom{{\sim}}$&  [\%]$^c$ \\

		\midrule
		 Multi-                            & \multirow{2}{*}{S} & \multirow{2}{*}{1}&\multirow{2}{*}{\xmark}&\multirow{2}{*}{\xmark}& \multirow{2}{*}{56}& \multirow{2}{*}{47$\phantom{{\sim}}$}& \multirow{2}{*}{$\phantom{-1}3.2$}  \\
          scale 2D~\cite{cui2021asymmetric} &  & &&& &  &  \\
		 \midrule
 		\multirow{3}{*}{Ctxformer~\cite{koyuncu2022contextformer}}& S& 1& \xmark& \xmark& && $\phantom{1}\!\!-5.0$\\
 		& S& 3& \xmark& \xmark& \multicolumn{2}{c|}{${\sim}36\cdot10^3$}& $\phantom{1}\!\!-6.9$\\
 		& S& 3& \cmark& \xmark& && $-10.2$\\
 		\midrule
 		pCtxtformer&\multirow{2}{*}{P}& \multirow{2}{*}{3}& \multirow{2}{*}{\xmark}& \multirow{2}{*}{\xmark}& \multirow{2}{*}{1121}& \multirow{2}{*}{$1112\phantom{{\sim}}$} & \multirow{2}{*}{$\phantom{1}\!\!-3.5$}\\
 	 (\cref{sec_2})& & & & & &  & \\
 		\midrule
 		& P& 1& \xmark& \xmark& 242 & 233$\phantom{{\sim}}$& $\phantom{1}\!\!-3.6$\\
 		eCtxformer& P& 3& \xmark& \xmark& 253&244$\phantom{{\sim}}$& $\phantom{1}\!\!-5.2$\\
 		(optimized)$^d$& P& 3& \xmark& \cmark& {${\sim}20\cdot10^3$}&244$\phantom{{\sim}}$& $-12.6$\\
 		& P& 3& \cmark& \xmark& 253&244$\phantom{{\sim}}$& $-10.9$\\
 		& P& 3& \cmark& \cmark& {${\sim}20\cdot10^3$} &244$\phantom{{\sim}}$& $-14.7$\\
		\bottomrule
	\end{tabular}
	   % \addtolength{\tabcolsep}{-4pt}
\begin{tablenotes}
\item[$a$] S and P stand for parallel and serial context models.
\item[$b$] The entropy model complexity during encoding and decoding.
\item[$c$] Computed w.r.t. PSNR over VTM 16.2~\cite{jvet2019versatile} (intra). Lower means higher bitrate savings.
\item[$d$] The combination of the proposed EGR, SFG and Caching optimizations are applied.
\end{tablenotes}
\end{threeparttable}
\end{table}

\begin{figure*}[!t]
\centering
\includegraphics[width=1\linewidth]{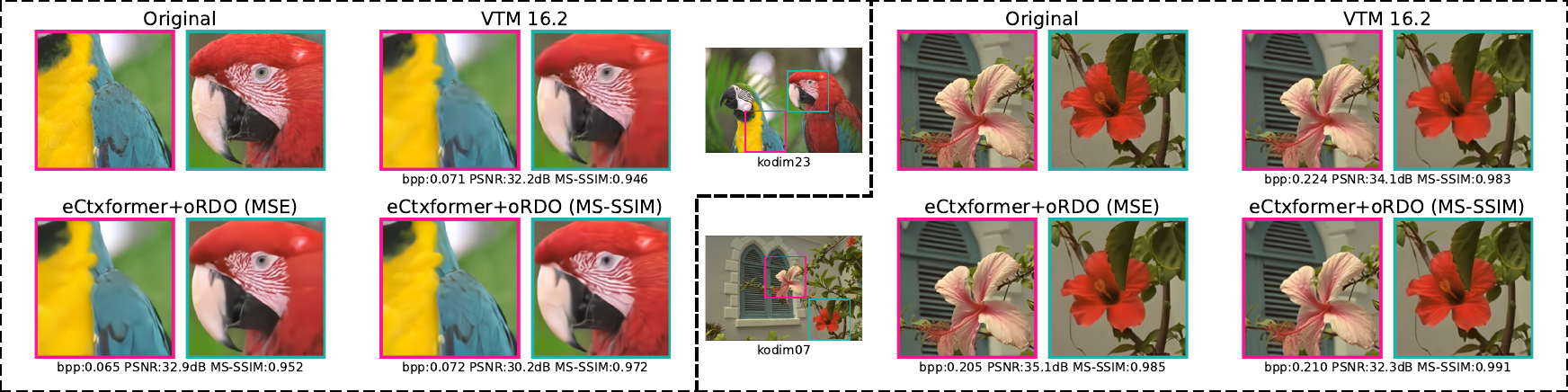}\hfill
\caption{Visual comparison of reconstructed images \textit{kodim23} (left) and \textit{kodim07} (right) from the Kodak image dataset. The images are compressed for the target bpp of 0.07 and 0.21, respectively. Both of our models, optimized for MSE and MS-SSIM, produce visually and objectively higher quality images for a lower bit-stream size than VTM 16.2~\cite{jvet2019versatile}.}
\label{fig_viz}
\end{figure*}
\begin{table*}[!t] % <-- HERE
	\centering
	\caption{The architecture of the proposed model using the eContextformer. Each row describes one layer or component of the model. ``Conv: $K_c{\times}K_c{\times}N\,\text{s}2$'' is a convolutional layer with kernel size of $K_c{\times}K_c$, number of $N$ output channels with a stride ``s'' of 2. Similarly, ``Deconv'' stands for transposed convolutions. ``Dense'' layers are specified by their output dimension, whereas $D_1\,{=}\,2M \,{+}\, d_e$, and $D_2\,{=}\,(4k_{m}M)/N_{cs}$.	
	}
	\label{tab:model_description}

\begin{tabular}{@{}cccccc@{}}
		\toprule[0.8pt]
		Encoder & Decoder & Hyperencoder & Hyperdecoder& Context Model & Entropy Parameters \\
		\midrule[0.4pt]
		Conv: $3{\times}3{\times}N\;{\text{s}}2$ & $2{\times}$ResBlock: $3{\times}3{\times}M$ & Conv: $5{\times}5{\times}N\;{\text{s}}1$ &  Conv: $5{\times}5{\times}M \;{\text{s}}1$ & eContextformer: & Dense: $(2D_1{+}D_2)/3$ \\
		
		GDN & Deconv: $3{\times}3{\times}N\; {\text{s}2}$ & Leaky ReLU & Deconv: $5{\times}5{\times}M \;{{\text{s}}2}$ 		&$\{L,K,d_e,\mathit{d}_{mlp},h,N_{cs}\}$& GELU\\
		
		Attention Module & iGDN & Conv: $5{\times}5{\times}N \;{\text{s}}1$& Leaky ReLU && Dense: $(D_1{+}2D_2)/3$\\
		
		Conv: $3{\times}3{\times}N\;{\text{s}}2$ & Deconv: $3{\times}3{\times}N\;{{\text{s}}2}$ & Conv: $5{\times}5{\times}N\;{\text{s}}2$ & Conv: $5{\times}5{\times}M \;{\text{s}}1$ && GELU\\
		
		GDN & iGDN & Leaky ReLU& Deconv: $5{\times}5{\times}\frac{3}{2}M \;{{\text{s}}2}$&& Dense: $D_2$ \\
		
		Conv: $3{\times}3{\times}N\;{\text{s}}2$ & Deconv: $3{\times}3{\times}N\;{{\text{s}}2}$ & Conv: $5{\times}5{\times}N\;{\text{s}}1$ & Leaky ReLU \\
		
		GDN & Attention Module & Conv: $5{\times}5{\times}N\;{\text{s}}2$ &Deconv: $5{\times}5{\times}2M \;{{\text{s}}2}$\\
		
		Conv: $3{\times}3{\times}N\;{\text{s}}2$ & iGDN & & Leaky ReLU\\
		
		Conv: $1{\times}1{\times}M\;{\text{s}}2$ & Deconv: $3{\times}3{\times}3\;{{\text{s}}2}$ & & \\
		\bottomrule[0.8pt]
	\end{tabular}
\end{table*}
\subsection{Visual Quality}
\cref{fig_viz} allows for a visual quality comparison between our models with oRDO applied, and VTM 16.2~\cite{jvet2019versatile}. The figure shows enlarged crops from two images from the Kodak dataset -- \textit{kodim07} and \textit{kodim23}. We compressed the images using each algorithm and targeting the same bitrate. In the figure, VTM 16.2~\cite{jvet2019versatile} exhibit noticeable artifacts such as smear and aliasing. On the other hand, our models offer superior visual quality compared to VTM 16.2~\cite{jvet2019versatile}, and are better at preserving contours and high-frequency details. According to our experience, the MSE-optimized model  excels in producing sharper edges, while the MS-SSIM optimized one is better at preserving texture grain.

\section{Conclusion}
This work introduces eContextformer -- an efficient and fast upgrade to the Contextformer. We conduct extensive experimentation to reach a fast and low-complexity context model while presenting state-of-the-art results. Notably, the algorithmic optimizations we provide further reduce the complexity by 84\%. Aiming to close the amortization gap, we also experimented with an encoder-side iterative algorithm. It further improves the rate-distortion performance and still has lower complexity than the state-of-art video compression standard. Undoubtedly, there are more advanced compression algorithms yet to be discovered which employ better non-linear transforms and provide more energy-compacted latent space. This work focuses on providing an efficient context model architecture, and defer such an improved transforms to future work.

\bibliographystyle{IEEEtran}
\bibliography{bib}
% \vskip -2\baselineskip plus -1fil
\begin{IEEEbiography}[{\includegraphics[width=1in,height=1.25in,clip,keepaspectratio]{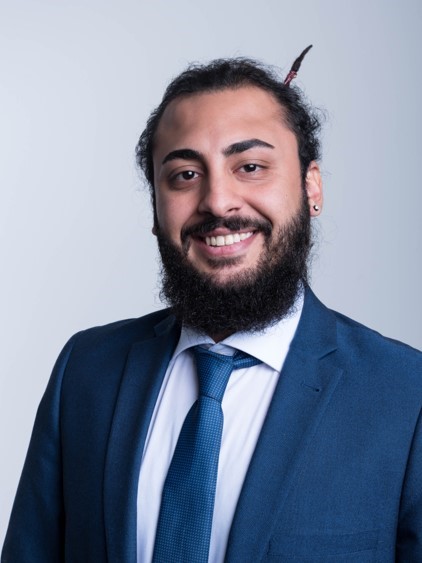}}]{A. Burakhan Koyuncu}
received the B.Sc. and M.Sc. (high distinction) degrees in electrical and computer engineering from the Technical University of Munich (TUM) in 2016 and 2019, respectively. During his bachelor’s, he worked on methods for neuromorphic architectures and bio-inspired artificial neural networks, and during his master’s, he researched deep learning-based control algorithms for driving simulators. Since 2020, he is pursuing the Ph.D. degree at the Chair of Media Technology at TUM and the Audiovisual Technology Laboratory at Huawei Technologies, Munich. Since 2021, he has been actively involved in the development of the JPEG AI standard that was issued by ISO/IEC. His research focuses on deep learning-based image and video processing, compression and coding.
\end{IEEEbiography}
\vskip -2\baselineskip plus -1fil
\begin{IEEEbiography}[{\includegraphics[width=1in,height=1.25in,clip,keepaspectratio]{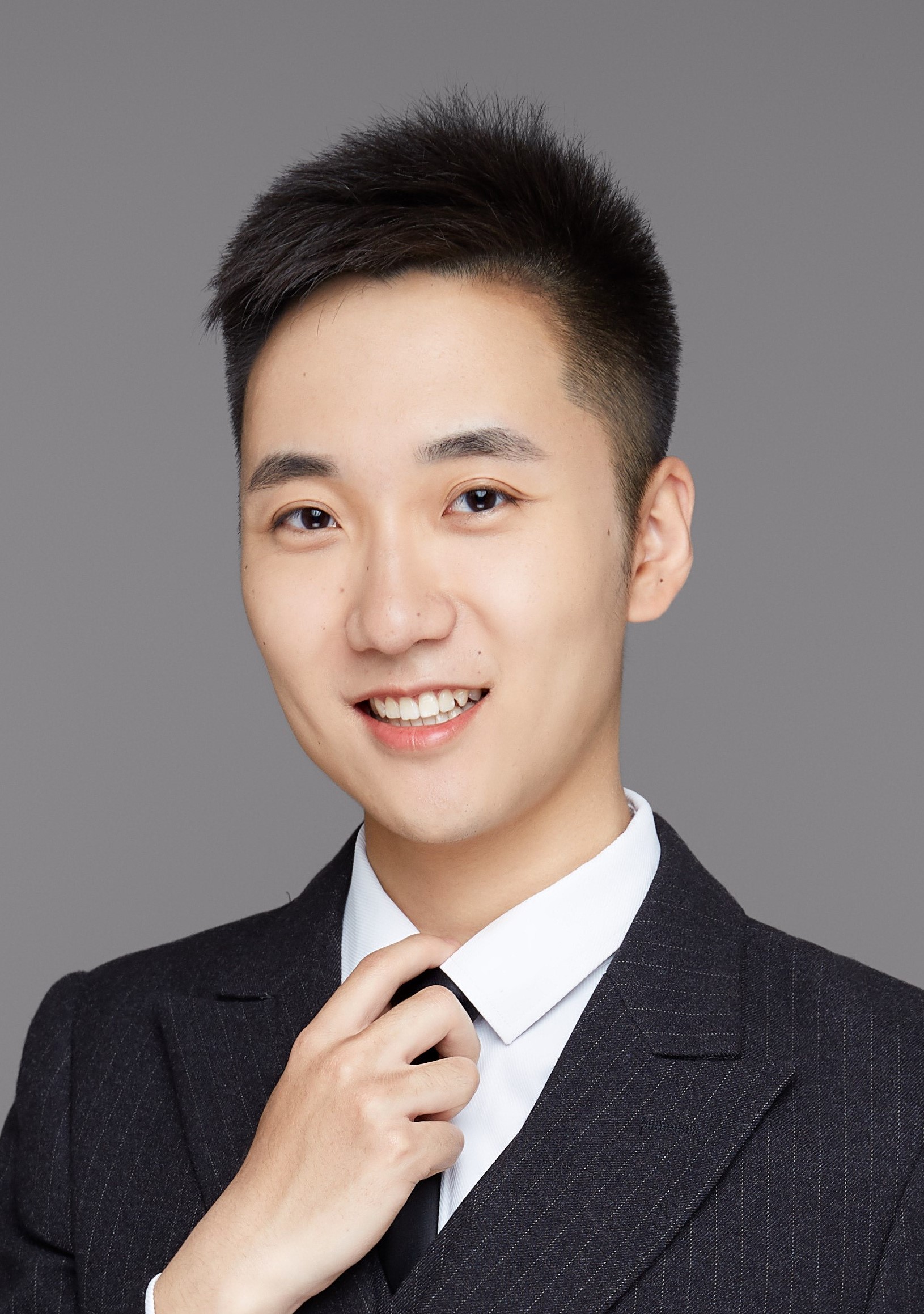}}]{Panqi Jia} obtained his master's degree in electrical engineering from Leibniz Universität Hannover (LUH) in Hannover, Germany in 2019. During his bachelor's degree, he worked on the application of Ultra-wideband radar. During his master's degree, he researched Car to X (C2X) communication systems and stereo matching algorithms in the vehicle communication area. Since 2020, he has been a joint PhD student at Huawei Technologies, which has collaborated with the Chair of Multimedia Communications and Signal Processing at Friedrich Alexander Universität (FAU). He conducts research on methods for image and video compression and deep learning.
\end{IEEEbiography}
\vskip -2\baselineskip plus -1fil
\begin{IEEEbiography}[{\includegraphics[width=1in,height=1.25in,clip,keepaspectratio]{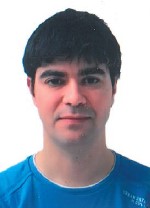}}]{Atanas Boev}
is a researcher with expertise in video compression, lightfield displays and human stereopsis. He defended his PhD in Tampere University of Technology, Finland in 2012. In 2013 he was visiting researcher at Holografika KFT, Hungary, doing development and implementation of lightfield rendering algorithms. In 2014 he was post-doctoral researcher in Tampere University of Technology working on modern signal processing methods for lightfield displays. Currently, he is a principal engineer in the Huawei European Research Centre, working on AI video compression, color perception and HDR tone mapping.
\end{IEEEbiography}
\vskip -2\baselineskip plus -1fil
\begin{IEEEbiography}[{\includegraphics[width=1in,height=1.25in,clip,keepaspectratio]{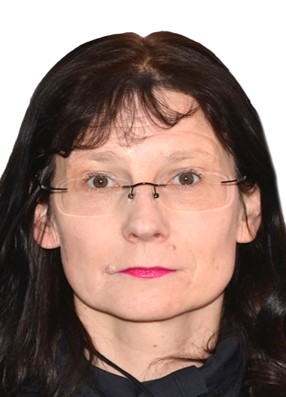}}]{Elena Alshina}
graduated from Moscow State University, got her PhD in Mathematical Modelling in 1998.
She (together with Alexander Alshin) got the Gold medal of the Russian Academy of Science in 2003 for series of research papers on Computer Science. In 2006 she joined Samsung Electronics and was part of the team that submitted the top-performing response for HEVC/H.265 Call for Proposals in 2010. Since that time she is an active participant in international video codec standardization. She was chairing multiple core experiments and AhGs in JCTVC, JVET, MPEG. Since 2018 she is Chief Video Scientist in Huawei Technologies (Munich), also the director for Media Codec and Audiovisual technology Labs. She has authored several scientific books, more than 100 papers in various journals and international conferences, 200+ patents. She currently is a co-chair of JVET Exploration on Neural Network-based video coding and JPEG AI standardization project co-chair and co-editor. Her current research interests include AI-based image and video coding, signal processing, computer vision.\end{IEEEbiography}
\vskip -2\baselineskip plus -1fil
\begin{IEEEbiography}[{\includegraphics[width=1in,height=1.25in,clip,keepaspectratio]{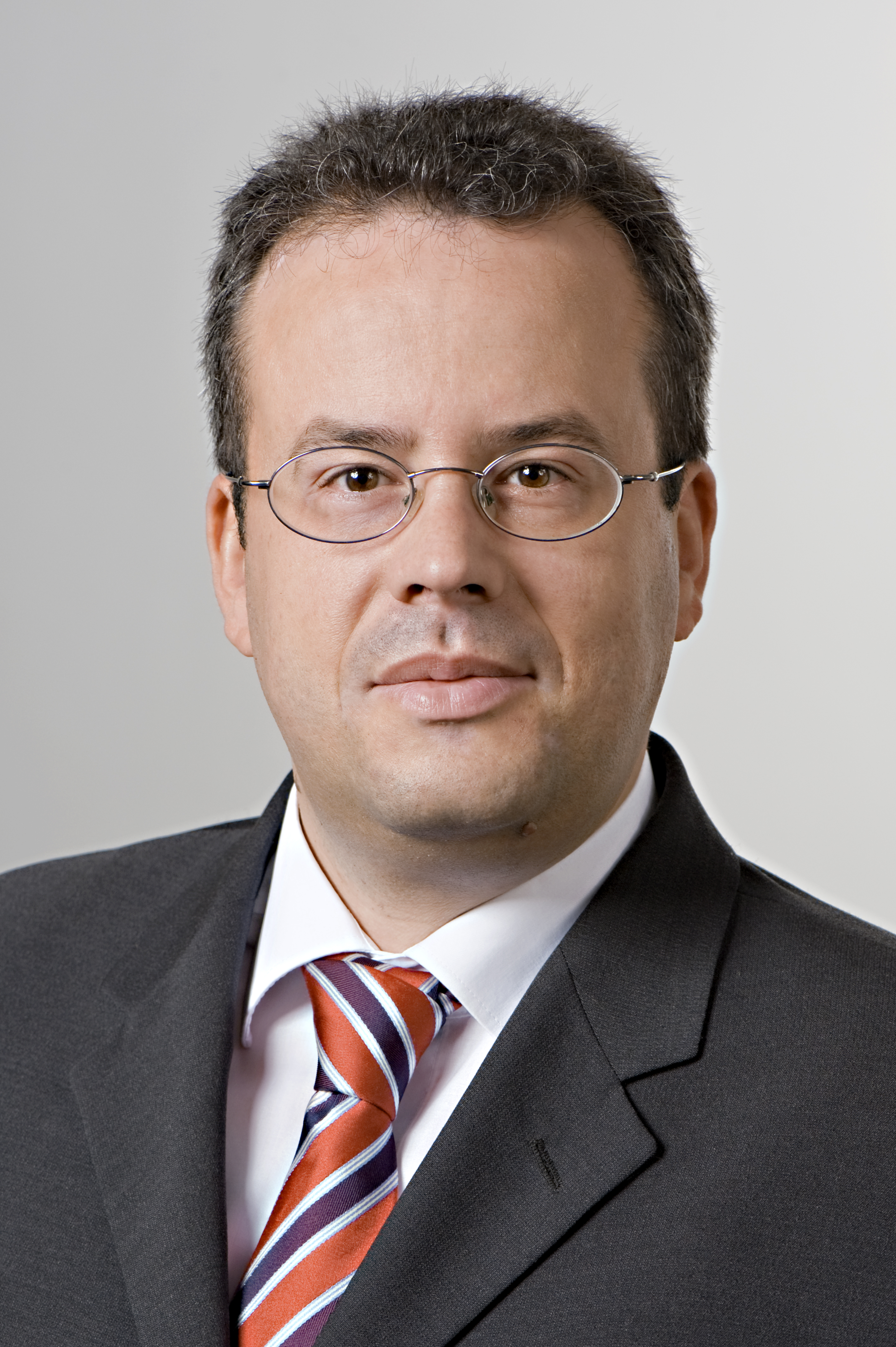}}]{Eckehard Steinbach} (IEEE M’96, SM’08, F’15) received the Ph.D. degree in electrical engineering from the University of Erlangen–Nuremberg, Germany, in 1999. In February 2002, he joined the Department of Computer Engineering, Technical University of Munich (TUM), where he is a Professor of media technology. His research interests are in the area of visual-haptic information processing and communication, telepresence and teleoperation, and networked and interactive multimedia systems.
\end{IEEEbiography}
\end{document}